\documentclass[12pt]{article}

\usepackage{amsmath}
\usepackage{amsfonts}
\usepackage{amssymb}
\usepackage{graphicx}
\usepackage{psfrag}
\usepackage{graphicx}
\usepackage{psfrag}

\newcommand{\be}{\begin{equation}}
\newcommand{\ee}{\end{equation}}
\newcommand{\ba}{\begin{eqnarray}}
\newcommand{\ea}{\end{eqnarray}}

\begin{document}
\begin{center}
{\bf{\large A New Class of Solvable Two-dimensional Scalar Potentials for Graphene}}\\
\vspace{1cm}
M. V. Iof\/fe$^{1,}$\footnote{E-mail: ioffe000@gmail.com; corresponding author},
D. N. Nishnianidze$^{2,}$\footnote{E-mail: cutaisi@yahoo.com}\\
\vspace{0.5cm}
$^1$ Saint Petersburg State University, 7/9 Universitetskaya nab., St.Petersburg, 199034 Russia.\\
$^2$ Akaki Tsereteli State University, 4600 Kutaisi, Georgia.\\

\end{center}
\vspace{0.5cm}
\hspace*{0.5in}
\vspace{1cm}
\hspace*{0.5in}
\begin{minipage}{5.0in}
{\small
In the present paper, a systematic approach is presented for solution of two-dimensional massless Dirac equation with external electrostatic potential applied. This approach is based on the new - asymmetric - form of SUSY-like intertwining relations. It allows to build a wide variety of pairs of SUSY-partner external scalar potentials. If one of them is simple enough to be solvable, its partner is also solvable although it may have a non-trivial dependency on both coordinates. Physically, this kind of problems is related to the description of graphene and some other materials with external potential. Solvability obtained by means of asymmetric form of SUSY intertwining relations allows to extend the class of analytically solvable two-dimensional models.
}
\end{minipage}


\section{Introduction.}

During last years, analytical study of two-dimensional massless Dirac equation with external scalar potentials was motivated by its connection to the theory of graphene and some other materials \cite{novoselov} -- \cite{novoselov-4}. In this context, the two-components of wave functions of Dirac equation describe the probability distribution of electron carriers and correspond to two different sublattices in graphene. For the first sight, the two-dimensional Dirac equation seems to be not too hard problem to solve, but for non-trivial potentials it is not so simple due to matrix character of equation. Analytical solvability of the problem was demonstrated by using different approaches for a very few potentials, only. In particular, it was done for potentials without dependence on one of coordinates or for the simple dependence on it \cite{peres} -- \cite{peres-12}.

Among different approaches to the problem, the method of supersymmetrical (SUSY) Quantum Mechanics \cite{witten} -- \cite{matveev} was applied effectively \cite{susy} -- \cite{samsonov} for building of such potential that the corresponding Dirac equation is solvable. Shortly speaking, these potentials are the superpartners of more simple potentials for which solutions of Dirac equations can be found more or less straightforwardly. Superpartnership of two Dirac operators means that they are intertwined by some "supercharge" operator, and their wave functions are connected to each other. These intertwining relations represent the most important ingredient of nonrelativistic supersymmetrical algebra.

Recently, the generalization of supersymmetrical intertwining relations was proposed in the mathematical context in papers \cite{shemyakova}, \cite{shemyakova-2}. In contrast to the usual SUSY intertwining, this generalization includes different supercharge operators in the l.h.s. and r.h.s. of the intertwining relations. Just by this reason, we call it {\it asymmetric.} Such kind of generalization is not suitable for the usual stationary Schr\"odinger spectral problem but it was shown to be useful for some other problems where no spectral parameter must be found. Examples of such problems are massless Dirac equation \cite{INP}, \cite{jakubsky} and Fokker-Planck equation \cite{fokker} .

The structure of the paper is the following. In Section 2, the asymmetric intertwining relations for two-dimensional massless Dirac equation with scalar external potential are introduced. In particular, it is demonstrated why the standard SUSY intertwining does not work for this problem. In Section 3, this intertwining is analyzed for the specific anzats, and the problem is reduced to the system of three nonlinear differential equations. This system is not solvable in a general form, and in Section 4, the particular class of problems with one of partner potentials $V_2(x_2)$ depending only on one of coordinates is investigated. It is shown that three different options appeared here, and in all of them, the partner potential $V_1(\vec x)$ depends non-trivially on both coordinates. A few specific examples of $V_2(x_2)$ are studied in Section 5 where non-trivial two-dimensional partner potentials are found, and corresponding wave functions of different kinds, including zero modes of supercharge, are calculated. Figures with presentation of obtained potentials for different choice of parameters are included in Appendix.

\section{Asymmetrical intertwining relations.}

In a standard Supersymmetrical Quantum Mechanics, interwining relations are the most important ingredients, they just provide the connection between spectra and wave functions for the pair of superpartner Schr\"odinger Hamiltonians. In the present case, we consider the pair of two-dimensional massless Dirac operators with some scalar potential:
\be
D_i=-i\sigma_1\partial_1 - i\sigma_2\partial_2 + V_i(\vec x); \quad i=1, 2; \quad \vec x=(x_1, x_2);\quad \partial_k\equiv \frac{\partial}{\partial x_k},
\label{dirac}
\ee
where $\sigma_i$ are Pauli matrices in the standard representation:
\be
\sigma_1=\left(
            \begin{array}{cc}
              0 & 1 \\
              1 & 0 \\
            \end{array}
          \right);\quad
          \sigma_2=\left(
            \begin{array}{cc}
              0 & -i \\
              +i & 0 \\
            \end{array}
          \right); \quad
           \sigma_3=\left(
            \begin{array}{cc}
              1 & 0 \\
              0 & -1 \\
            \end{array}
          \right).    \nonumber
\ee
We are interested in the corresponding two-component wave functions:
\be
D_i\Psi^{(i)}(\vec x)=0;\quad \Psi^{(i)}=\left(
            \begin{array}{c}
              \Psi^{(i)}_1(\vec x) \\
              \Psi^{(i)}_2(\vec x) \\
            \end{array}
          \right);.
\label{psi}
\ee
The task is to find analytically solutions of the Dirac equation with operator $D_1$ for the case that the partner $D_2$ is simple enough, so that its wave functions $\Psi^{(2)}(\vec x)$ can be found analytically. Such partnership is provided by the intertwining relations between $D_1$ and $D_2:$
\be
D_1\,\, N = M\,\, D_2,
\label{int}
\ee
where $N(\vec x),\, M(\vec x)$ are $2\times 2$ matrix differential operators of first order:
\be
N(\vec x) \equiv A_i\partial_i + A(\vec x);\quad M(\vec x) \equiv B_i\partial_i + B(\vec x),
\label{NM}
\ee
with summation over repeated index $i = 1,2.$ Here, coefficients $A_i,\, B_i$ are constant $2\times 2$ matrices while matrices $A(\vec x),\, B(\vec x)$, in general, depend on coordinates $\vec x.$ Here we should pay attention to the mismatch
 between "supercharge" operators $N$ and $M.$ This asymmetry is just what distinguishes intertwining (\ref{NM}) from the standard intertwining relations of SUSY Quantum Mechanics.

Now, the key task is to develop a regular scheme for solution of generalized intertwining relations (\ref{int}). It means, we have to find from (\ref{int}) both potentials
$V_i(\vec x),$ the constant matrices $A_i, B_i$ and $2\times 2$ matrix functions $A(\vec x), B(\vec x).$ For this purpose, Eq.(\ref{int}) must be decomposed over Pauli matrices and over different partial derivatives. This procedure gives:
\ba
&&\sigma_k A_n-B_k\sigma_n+\sigma_n A_k-B_n\sigma_k=0; \label{2}\\
&&i\sigma_k A(\vec x)-V_1(\vec x) A_k = -V_2(\vec x)B_k+iB(\vec x)\sigma_k; \label{3}\\
&&i\sigma_k(\partial_kA(\vec x))-V_1(\vec x) A(\vec x)=-B_k(\partial_kV_2(\vec x))-B(\vec x)V_2(\vec x). \label{4}
\ea
Eq(\ref{2}) gives relations between constant coefficients $a's:$
\ba
&&A_1=a_0^{(1)}+\vec{a}^{(1)}\cdot \vec{\sigma} ;\nonumber\\
&&A_2=a_0^{(2)}-a_2^{(1)}\cdot\sigma_1+a_1^{(1)}\cdot\sigma_2+a_3^{(2)}\cdot\sigma_3; \label{a}\\
&&B_1=a_0^{(1)}+a_1^{(1)}\cdot\sigma_1-a_2^{(1)}\cdot\sigma_2-a_3^{(1)}\cdot\sigma_3; \nonumber\\ &&B_2=a_0^{(2)}+a_2^{(1)}\cdot\sigma_1+a_1^{(1)}\cdot\sigma_2-a_3^{(2)}\cdot\sigma_3.
\label{b}
\ea

Before turning to the next steps of our procedure for solution of (\ref{int}), it is appropriate to clarify the situation for the case of ordinary {\it symmetric} intertwining relations  with a single operator $N=M,$ i.e. with $A_i = B_i$ and $A(\vec x) = B(\vec x).$ From (\ref{a}), (\ref{b}), it follows that $a_2^{(1)}=a_3^{(1)}=a_3^{(2)}=0,$ so that
$$
A_1=B_1=a_0^{(1)}+a_1^{(1)}\sigma_1;\quad A_2=B_2=a_0^{(2)}+a_1^{(1)}\sigma_2.
$$
Then, Eq.(\ref{3}) can be rewritten as:
$$
i[\sigma_1, A(\vec x)]=(V_1(\vec x)-V_2(\vec x)) A_1;\quad
i[\sigma_2, A(\vec x)]=(V_1(\vec x)-V_2(\vec x)) A_2.
$$
Together with expressions above for $A_1=B_1$ and $A_2=B_2,$ we obtain that all other coefficients $a's$ also vanish:
$$
a_0^{(1)}=a_0^{(2)}=a_1^{(1)}=0.
$$
Thus, no solutions of (\ref{int}) for the {\it symmetric} case exist.

\section{The ansatz $A_1=1,\, A_2=0$.}

Let us now return to the general, {\it asymmetric,} form of intertwining. Starting from this point, we choose the anzats for the matrices $A_1=1,\,\, A_2=0,$ which induces $B_1=1,\,\, B_2=0,$ according to (\ref{int}). This anzats seems to be more interesting than the one used in \cite{INP}.
Now, Eq.(\ref{3}) gives:
\ba
V_1(\vec x)-V_2(\vec x)&=&i(\sigma_1 A(\vec x)-B(\vec x)\sigma_1); \nonumber
\\
\sigma_2A(\vec x)&=&B(\vec x)\sigma_2.  \nonumber
\ea
In terms of matrix coefficients, this means:
\be
 A(\vec x)=\left(
            \begin{array}{cc}
              a_{11} & a_{12} \\
              a_{21} & a_{22} \\
            \end{array}
          \right); \quad
          B(\vec x)=\left(
            \begin{array}{cc}
              a_{22} & -a_{21} \\
              -a_{12} & a_{11} \\
            \end{array}
          \right),  \nonumber
\ee
and
\be
V_1(\vec x)-V_2(\vec x)=2i \left(
            \begin{array}{cc}
              a_{21} & 0 \\
              0 & a_{12} \\
            \end{array}
          \right). \nonumber
\ee
Since the case of scalar potentials $V_i(\vec x)$ is considered, the diagonal elements have to coincide:
\ba
&&a_{12}(\vec x)=a_{21}(\vec x)\equiv -if(\vec x); \nonumber \\
&&V_1(\vec x)-V_2(\vec x)=2f(\vec x),
\label{4.1}
\ea
where $f(\vec x)$ is still an arbitrary real function.

Comparing (\ref{4}) with derivative of (\ref{3}) over $\partial_k$ and summation over $k$, one obtain:
\be
(\partial_k V_1(\vec x))A_k +i(\partial_k B(\vec x))\sigma_k=V_1(\vec x)A(\vec x)-V_2(\vec x)B(\vec x);   \nonumber
\ee
\be
(\partial_1V_1)+i\left(
            \begin{array}{cc}
              -\partial_1a_{12} & \partial_1a_{22}\\
              \partial_1a_{11} & -\partial_1a_{12} \\
            \end{array}
          \right) +
          \left(
            \begin{array}{cc}
              \partial_2a_{12} & \partial_2a_{22} \\
              -\partial_2a_{11} & -\partial_2a_{12} \\
            \end{array}
          \right)= \left(
            \begin{array}{cc}
              V_1a_{11}-V_2a_{22} & V_1a_{12}+V_2a_{21} \\
              V_1a_{21}+V_2a_{12} & V_1a_{22}-V_2a_{11} \\
            \end{array}
          \right). \nonumber
\ee
Therefore, diagonal elements of $A(\vec x)$ are expressed in terms of an arbitrary function $g(\vec x):$
\be
a_{11}=(\partial_1-i\partial_2)g(\vec x);\quad
a_{22}=(\partial_1+i\partial_2)g(\vec x), \nonumber
\ee
and the system of three nonlinear equations for unknown functions $f(\vec x),\, g(\vec x),\, V_2(\vec x)$ must be fulfilled:
\ba
(\partial_1^2+\partial_2^2)g(\vec x)&=&-2f(\vec x)(V_2(\vec x)+f(\vec x));\label{5.3}\\
\partial_1(V_2(\vec x)+f(\vec x))&=&2f(\vec x)\partial_1g(\vec x);\label{5.4}\\
\partial_2f(\vec x)&=&2(V_2(\vec x)+f(\vec x))\partial_2g(\vec x). \label{5.5}
\ea

The solutions $f(\vec x), \, g(\vec x),$ if obtained from (\ref{5.3}) - (\ref{5.4}), will be used below to construct intertwining operators:
\ba
N(\vec x)&=&\partial_1 +\left(
            \begin{array}{cc}
              (\partial_1-i\partial_2)g(\vec x) & -if(\vec x) \\
              -if(\vec x) & (\partial_1+i\partial_2)g(\vec x) \\
            \end{array}
          \right);\label{NNN}\\
M(\vec x)&=&\partial_1 +\left(
            \begin{array}{cc}
              (\partial_1+i\partial_2)g(\vec x) & if(\vec x) \\
              if(\vec x) & (\partial_1-i\partial_2)g(\vec x) \\
            \end{array}
          \right).
\label{NMNM}
\ea

It is the system of equations (\ref{5.3}) - (\ref{5.5}) that is becoming the main object of research now. Derivation of its general solution seems to be too difficult task. But limiting ourselves to one or another choice of ansatz for which the initial potential $V_2(\vec x)$ is rather simple, we get a chance to find nontrivial interesting solutions for its partner. The simplest case when $V_2$ is a constant was studied in \cite{INP} and \cite{jakubsky}.
Here, we will consider again this case more completely and systematically. Also, we study the more general class of systems with $V_2$ actually depending on one of coordinates $V_2=V_2(x_2),$ while the partner potential $V_1(\vec x)$ depends on both coordinates. It is obvious that the case of $V_2(x_1)$ can be described similarly.

If the problem with potential $V_2$ is solvable, and if its wave functions $\Psi^{(2)}(\vec x)$ are known, we may use intertwining relations (\ref{int}) to find the wave functions $\Psi^{(1)}(\vec x)$ of the partner potential $V_1(\vec x)$ as $\Psi^{(1)}=N\Psi^{(2)}$. Also, it is well known that for some SUSY Quantum Mechanical systems when intertwining is present but Hamiltonians are not factorizable (see \cite{zero-1} -- \cite{zero-4}) an additional class of wave functions exists. We mean so called zero modes $\Omega$ of supercharges. In our present case of Dirac equation and asymmetric intertwining (\ref{int}), equation $D_1\Psi^{(1)}(\vec x)=0$ may have also an additional class of solutions besides those obtained by $\Psi^{(1)}=N\Psi^{(2)}.$ Such solutions, just the zero modes $\Omega(\vec x)$, are the common solutions of two equations:
\be
M^{\dagger}((\vec x))\Omega(\vec x)=0;\quad D_1((\vec x))\Omega(\vec x)=0;\quad \Omega(\vec x)=(\Omega_1(\vec x),\, \Omega_2(\vec x))^T,
\label{0-mode-1}
\ee
where $M^{\dagger}$ is hermitian conjugated of (\ref{NMNM}).

\section{Solutions for $V_2=V_2(x_2)$.}

Dependence of $V_2$ only on one of coordinates simplifies solvability of the system (\ref{5.3}) - (\ref{5.5}). First, Eq.(\ref{5.4}) allows to express $f(\vec x)$ in terms of $g(\vec x)$ and arbitrary function $\varphi(\vec x_2):$
\be
f(\vec x)=\varphi(x_2)\exp{(2g(\vec x))}.
\label{fg}
\ee
Inserting the derivative of (\ref{fg}) into (\ref{5.5}) we obtain:
\be
\varphi^{\prime}(x_2)\exp{(2g(\vec x))} = 2V_2(x_2)\partial_2g(\vec x),
\label{gV}
\ee
and therefore,
\ba
\exp{(-2g(\vec x))} &=& g_1(x_1) + g_2(x_2);\label{aa}\\
g_2(x_2)\equiv -\int\frac{\varphi^{\prime}(x_2)}{V_2(x_2)}dx_2;\quad f(\vec x)&=&
\frac{\varphi(x_2)}{ g_1(x_1) + g_2(x_2)};\quad V_2(x_2)=-\frac{\varphi^{\prime}(x_2)}{g_2^{\prime}(x_2)}.
\label{Vg}
\ea
where $g_1(x_1)$ is also, as well as $\varphi(x_2)$, an arbitrary function.

Thus, Eqs.(\ref{5.4}) and (\ref{5.5}) allowed to express functions $V_2(x_2),\, f(\vec x)$ and $g(\vec x)$ in terms of three arbitrary functions $\varphi(x_2), g_1(x_1)$ and $g_2(x_2).$ It remains now to solve the last still unsolved equation of the system above, namely, Eq.(\ref{5.3}). Calculating derivatives of (\ref{aa}) and using formulas (\ref{Vg}), we obtain:
\be
g_1^{\prime\prime}g_2+g_2^{\prime\prime}g_1+\frac{2(\varphi^2)^{\prime}g_1}{g_2^{\prime}}=
(g_1^{\prime})^2-g_1^{\prime\prime}g_1+(g_2^{\prime})^2-g_2^{\prime\prime}g_2+4\varphi^2-\frac{2(\varphi^2)^{\prime}g_2}{g_2^{\prime}}.
\label{3.22}
\ee
Action by $\partial_1\partial_2$ onto (\ref{3.22}) allows to separate variables:
\be
\frac{g_1^{\prime\prime\prime}}{g_1^{\prime}}+\frac{g_2^{\prime\prime\prime}+2(\frac{(\varphi^2)^{\prime}}{g_2^{\prime}})^{\prime}}{g_2^{\prime}}=0. \nonumber
\ee
Now, we have to solve two ordinary differential equations:
\ba
&&g_1^{\prime\prime}+\lambda^2g_1=c_1
\label{3.44}\\
&&4\varphi^2=-(g_2^{\prime})^2+\lambda^2g_2^2+2c_2g_2+c_3,
\label{3.444}
\ea
where $\lambda^2$ is a separation constant. A part of arbitrary real constants are fixed after substitution of (\ref{3.44}), (\ref{3.444}) back into (\ref{3.22}) (this check is necessary because of taking derivatives of (\ref{3.22}) above). Also, the condition that $g_1$ and $g_2$ are real functions must be taken into account.

In summary, the above analysis leads to three different solutions depending on the sign of separation constant $\lambda^2:$
\begin{description}
  \item[Solution I.]
  \ba
g_1(x_1)&=&\alpha_1e^{i\lambda x_1}+\beta_1e^{-i\lambda x_1}
\label{6.2}\\
4\varphi^2(x_2)&=&-(g_2^{\prime}(x_2))^2+\lambda^2g_2^2(x_2)-4\lambda^2\alpha_1\beta_1.
\label{66.2}
\ea
where $\lambda$ is an arbitrary non-vanishing real constant, and coefficients $\alpha_1,\, \beta_1$ are arbitrary mutually conjugated complex  $\alpha_1=\beta_1^{\star}.$ Using translation of coordinates, we can achieve the compact form:
\be
g_1(x_1)=a \cos(\lambda x_1),
\label{7.1}
\ee
and
\be
4\varphi^2=-(g_2^{\prime}(x_2))^2+\lambda^2g_2^2(x_2)-\lambda^2a^2.
\label{7.2}
\ee
Correspondingly, $g_2(x_2)$ is here an arbitrary real function, and potential $V_2(x_2)$ is expressed in terms of this function according to Eqs.(\ref{Vg}), (\ref{66.2}):
\be
V_2(x_2)=\frac{g_2^{\prime\prime}(x_2)-\lambda^2g_2(x_2)}{2\sqrt{-(g_2^{\prime}(x_2))^2+
\lambda^2g_2^2(x_2)-\lambda^2a^2}}.
\label{V2-1}
\ee

  \item[Solution II.]

 For the negative value of $\lambda^2$, replacement of $\lambda$ by $i\lambda$ and again, a suitable shift of coordinates, gives:
\ba
g_1(x_1)&=&a \cosh(\lambda x_1),
\label{7.11}\\
4\varphi^2&=&-g_2^{\prime}(x_2))^2-\lambda^2g_2^2(x_2)+\lambda^2a^2
\label{7.111}
\ea
and
\be
V_2(x_2)=\frac{g_2^{\prime\prime}(x_2)+
\lambda^2g_2(x_2)}{2\sqrt{-(g_2^{\prime}(x_2))^2-
\lambda^2g_2^2(x_2)+\lambda^2a^2}},
\label{V2-2}
\ee
where new $\lambda $ is again arbitrary real.

  \item[Solution III.]

  For the case $\lambda =0,$ function $g_1(x_1)$ admits a simpler form, and $g_2(x_2)$ is still arbitrary real function:
\ba
g_1(x_1)&=&\frac{1}{2}cx_1^2;  \label{6.3}\\
4\varphi^2(x_2)&=&-(g_2^{\prime}(x_2))^2+2c g_2(x_2),
\label{66.3}
\ea
with arbitrary real $c.$ In this case, the potential $V_2(x_2)$ is:
\be
V_2(x_2)=
\frac{g_2^{\prime\prime}(x_2)-
c}{2\sqrt{-(g_2^{\prime}(x_2))^2+2cg_2(x_2)}}.
\label{V2-3}
\ee
\end{description}

\section{Specific forms of potential $V_2(x_2)$.}

A class of potentials $V_2(x_2)$ above is evidently very wide. Indeed, the function $\varphi(x_2)$ is defined by an arbitrary function $g_2(x_2)$ (see Eq.(\ref{3.444})). After that, this chosen function $g_2(x_2)$ and obtained $\varphi(x_2)$ lead to the potential $V_2(x_2)$ according to relations either (\ref{7.2}), or (\ref{7.111}), or (\ref{66.3}). Below, we shall also consider an inverse problem, i.e. to look for functions $g_2(x_2)$ which lead to a given (rather simple) potential $V_2(x_2).$

\subsection{Models with $V_2(x_2)=0$}

As it is seen from (\ref{Vg}), just this exceptional case is impossible to analyse by the general formulas of the previous Section.
The initial system of equations (\ref{5.3}) - (\ref{5.5}) can be solved individually step by step in this special case (see also Example 1 in \cite{INP}, where another anzats for $N, M$ was chosen). Eqs.(\ref{5.4}), (\ref{5.5}) provide the connection between $f(\vec x)$ and $g(\vec x):$
\be
f(\vec x)=\gamma \exp{(2g(\vec x))};\quad \gamma=const.
\label{0-1}
\ee
In turn, Eq.(\ref{5.3}) is now equation for the only function $g(\vec x):$
\be
(\partial_1^2+\partial_2^2)g(\vec x)=-2\gamma^2\exp{(4g(\vec x))}.
\label{0-2}
\ee
This nonlinear equation occurs within the framework of burning theory \cite{polyanin} (see also \cite{INP}), and two different solutions are known:
\ba
g^{(a)}(\vec x)&=&-\frac{1}{2}\ln{\cosh(Ax_1+Bx_2+C)}+\frac{1}{4}\ln{\frac{A^2+B^2}{4\gamma^2}};
\label{0-3}\\
g^{(b)}(\vec x)&=&-\frac{1}{2}\ln[(x_1+A)^2+(x_2+B)^2+C]+\frac{1}{4}\ln(\frac{C}{\gamma^2}),
\label{0-4}
\ea
with arbitrary constants $A, B, C.$ It is clear that by the suitable shifts of coordinates, some of these constants can be fixed to simplify expressions without any loss of generality. Namely, we shall take $C=0$ in (\ref{0-3}) and $A=B=0$ in (\ref{0-4}). Thus, the partner potentials of $V_2(x_2)=0$ for these two solutions are, correspondingly:
\ba
V_1^{(a)}(\vec x)&=& \pm\frac{\sqrt{A^2+B^2}}{\cosh(Ax_1+Bx_2)};
\label{0-5}\\
V_1^{(b)}(\vec x)&=& \pm\frac{2\sqrt{C}}{x_1^2+x_2^2+C},
\label{0-6}
\ea
where signs $\pm$ equal $\gamma /|\gamma|.$

The two-component wave functions for initial operator $D_2$ with vanishing potential are very simple:
\be
\Psi^{(2)}_1(\vec x)=\psi_1(z);\quad
\Psi^{(2)}_2(\vec x)=\psi_2(\bar z); \quad z=x_1+ix_2;\,
\quad \bar z=x_1-ix_2,
\label{z}
\ee
where $\psi_1, \psi_2$ are arbitrary functions of their arguments.
Now, the components of wave functions $\Psi^{(1)}=N\Psi^{(2)}$ (see Eq.(\ref{NNN})) for the partner operator $D_1$ are different for models $(a)$ and $(b)$ above:
\begin{itemize}
  \item (a)
\ba
\Psi^{(1), a}_1
(\vec x)&=&(\partial_1-
\frac{A-iB}{2}\tanh(Ax_1+Bx_2))\psi_1(z)\mp \frac{i\sqrt{A^2+B^2}}{\cosh(Ax_1+Bx_2)}\psi_2(\bar z)
\label{a-1}\\
\Psi^{(1), a}_2(\vec x)&=&(\partial_1-\frac{A+iB}{2}\tanh(Ax_1+Bx_2))\psi_2(\bar z)\mp\frac{i\sqrt{A^2+B^2}}{\cosh(Ax_1+Bx_2)}\psi_1(z)
\label{a-2}
\ea

  \item (b)
\ba
\Psi^{(1), b}_1(\vec x)&=&(\partial_1-\frac{\bar z +A-iB}{x_1^2+x_2^2+C})\psi_1(z)\mp \frac{2i\sqrt{C}}{x_1^2+x_2^2+C}\psi_2(\bar z)
\label{a-3}\\
\Psi^{(1), b}_2(\vec x)&=&(\partial_1-
\frac{z+A+iB}{x_1^2+x_2^2+C})\psi_2(\bar z)\mp \frac{2i\sqrt{C}}{x_1^2+x_2^2+C}\psi_1(z). \label{a-4}
\ea
\end{itemize}

Now, we have to build the wave functions of $D_1,$ which are simultaneously the zero modes of $M^{\dagger}.$ Straightforwardly, the system of equations (\ref{0-mode-1}) gives these components as:
\be
\Omega_1(\vec x)=\omega_1(\bar z)\exp(2g(\vec x));\quad \Omega_2(\vec x)=\omega_2(z)\exp(2g(\vec x));\quad z=x_1+ix_2;\, \bar z=x_1-ix_2,
\label{comp}
\ee
where each function $\omega_{1,2}$ depends on its own argument ($\bar z\,$ or $ z$), correspondingly, and in addition, satisfies second order differential equation:
\ba
&&\omega''_1(\bar z)+2[(\bar\partial^2g(\vec x))-2(\bar\partial g(\vec x))^2]\omega_1(\bar z)=0; \quad \bar\partial=\frac{1}{2}(\partial_1+i\partial_2);
\label{omega-1}\\
&&\omega''_2(z)+2[(\partial^2g(\vec x))-2(\partial g(\vec x))^2]\omega_2(z)=0; \quad \partial=\frac{1}{2}(\partial_1-i\partial_2).
\label{omega-2}
\ea
One can check that due to Eq.(\ref{5.3}), the expressions in square brackets of (\ref{omega-1}), (\ref{omega-2}) depend on $\bar z$ or $z,$ correspondingly. In both cases, $(a)$ and $(b)$ above, these expressions turn out to be equal to the constants, and Eqs.(\ref{omega-1}), (\ref{omega-2}) are easily solved. Thus, the dependence of zero modes (\ref{comp}) on coordinates for the models $(a)$ and $(b)$ is:
\ba
\Omega^{(a)}_1(\vec x)&=&
-i(A-iB)(\cosh(Ax_1+Bx_2))^{-1}[\beta^{(a)}\exp(\frac{A+iB}{2}\bar z)
-\alpha^{(a)}\exp(-\frac{A+iB}{2}\bar z)]; \label{O-1}\\
\Omega^{(a)}_2(\vec x)&=&
(\cosh(Ax_1+Bx_2))^{-1}
[\alpha^{(a)}\exp(\frac{A-iB}{2}z)+
\beta^{(a)}\exp(-\frac{A-iB}{2}z)]; \label{O-2}
\ea
and
\ba
\Omega^{(b)}_1(\vec x)=\frac{\alpha^{(b)}\bar z+i\beta^{(b)}\sqrt{C}}{x_1^2+x_2^2+C};
\label{O-3}\\
\Omega^{(b)}_2(\vec x)=\frac{\beta^{(b)}z+i\alpha^{(b)}\sqrt{C}}{x_1^2+x_2^2+C}
\label{O-4}
\ea
where $\alpha$'s and $\beta$'s are arbitrary constants.

\subsection{Models with $V_2(x_2)= m/2 =const >0.$}

Eq.(\ref{Vg}) gives in this case:
\be
\varphi(x_2)=-\frac{mg_2(x_2)+n}{2}; \, n=const,
\label{n}
\ee
Expressions for $g_2(x_2),$ $f(\vec x)$ and $V_1(\vec x)$ can be obtained (see (\ref{Vg})) for each of three options for $g_1(x_1)$ listed in the previous Section.

1) For the first, Eq.(\ref{7.2}) with (\ref{n}) is solvable, and two different solutions exist:
\ba
g_2^{\alpha}(x_2)&=&b\cos(kx_2)+ mA; \quad k^2\equiv m^2-\lambda^2>0;\quad A\equiv\pm\sqrt{\frac{b^2}{\lambda^2}+\frac{a^2}{m^2-\lambda^2}};
 \label{i}\\
g_2^{\beta}(x_2)&=&b\cosh(kx_2)+ mA;
\quad k^2\equiv \lambda^2-m^2>0, \label{ii}
\ea

According to (\ref{4.1}), the corresponding partner potentials are:
\ba
V_1^{\alpha}(\vec x)&=&\frac{m}{2}-
\frac{mb\cos(kx_2)+\lambda^2A}{a\cos(\lambda x_1)+b\cos(kx_2)+mA}; \quad k^2\equiv m^2-\lambda^2>0;
\label{V-i}\\
V_1^{\beta}(\vec x)&=&\frac{m}{2}-
\frac{mb\cosh(kx_2)+\lambda^2A}
{a\cos(\lambda x_1)+b\cosh(kx_2)+mA}; \quad k^2\equiv \lambda^2-m^2>0;
\label{V-ii}.
\ea
(both potentials have no singularities for arbitrary values of constants). Let us notice that while the $\alpha$-model is periodic on both variables, the potential of $\beta$-model is more interesting being periodic only on $x_1$ and constant asymptotically on $x_2.$ The Figure 1 represents dependence of potential (\ref{V-ii}) on parameter $A,$ which is changed from $A=0.0$ to $A=0.6$ with fixed $m=b=k=2.$ It is clear that the form and the scale of potential depend on $A$ rather weakly. In turn, the Figure 2 gives the dependence on parameter $k$ changing as $k=3,\, k=10, \, k=30$ with fixed $m=b=2,\, a=1.$ In this case, both the form and scale of potential depend strongly on parameter $k.$

It is necessary to stress that the constants $a,\,b,\,\lambda$ above are arbitrary real constants. For two different particular cases of special connections between these constants, our $\beta$-model (\ref{V-ii}) coincides with models $A$ and $B$ of the paper \cite{jakubsky}. Specifically, if one takes for our $\beta-$model (\ref{ii}), (\ref{V-ii}) a particular case with related constants $a = k^2\equiv (\lambda^2-m^2)$ and $b=\lambda^2,$ the potential (\ref{V-ii}) coincides exactly with the model (37) of \cite{jakubsky} after an additional suitable redefinition of our constants $m\to 2m;\, \lambda\to 2k;\, k\to 2\omega.$ The same latter redefinition of constants, together with a suitable shift of coordinates and special constraints between parameters reduce our $\beta-$model to the model (52) of \cite{jakubsky}. Thus, both potentials (37) and (52) of \cite{jakubsky} are some particular cases of our potential (\ref{V-ii}).

2) For the second, Eq.(\ref{7.111}) is also solvable but with only one solution:
\be
g_2(x_2)=b\cos(kx_2)+ mA;\quad  A\equiv\sqrt{\frac{a^2}{m^2-\lambda^2}-\frac{b^2}{\lambda^2}}; \label{i-2}
\ee
\be
V_1(\vec x)=\frac{m}{2}-
\frac{mb\cos(kx_2)-
\lambda^2A}{a\cosh(\lambda x_1)+ b\cos(kx_2)+mA};\,\,k^2\equiv m^2+\lambda^2 .
\label{V-i-2}
\ee

3) For the third, the solution of Eq.(\ref{66.3}) is given by:
\be
g_2(x_2)=B\cos(mx_2)+\gamma;\quad \gamma\equiv\frac{c^2+m^4B^2}{2cm^2},
\label{add}
\ee
if the constants are connected by
$$n=\frac{c^2-m^4B^2}{2mc}.$$
The partner potential is:
\be
V_1(\vec x)=\frac{m}{2}-
\frac{mB\cos(mx_2)+c/m}{\frac{1}{2} cx_1^2+B\cos(mx_2)+\gamma},
\label{V-i-3}
\ee
and its form is presented at Figure 3 for some values of parameters. It is evident that by choosing values of parameters, a wide variety of periodic row of holes can be obtained.

As in the previous case of Subsection 5.1, the wave functions $\Psi^{(2)}(\vec x)$ for the initial $V_2=m/2$ are easy to find directly from the Dirac equation. Indeed, the second component $\Psi^{(2)}_2$ is expressed via $\Psi^{(2)}_1,$ and in its turn $\Psi_1^{(2)}$ satisfies the second order differential equation, which in the present case reduces to the Helmholtz equation. Thus, the solutions have the form of plane waves:
\ba
\Psi^{(2)}_1(\vec x)&=& C \cdot\exp(i\vec q \vec x);\quad (\vec q)^2=\frac{m^2}{4}; \label{helm-1}\\
\Psi^{(2)}_2(\vec x)&=& C \cdot\frac{2i(iq_1-q_2)}{m} \exp(i\vec q \vec x),  \label{helm-2}
\ea
with arbitrary direction of two-dimensional momentum $\vec q.$

At this point, we have to note that the limit $m\to 0$ can not be performed in this Subsection.
The inequalities between parameters in relations (\ref{ii}) and later on can not be fulfilled. Also, the wave functions $\Psi^{(2)}_{1,2}$ in (\ref{helm-1}), (\ref{helm-2})) are not reduced to the holomorphic or anti-holomorphic functions (see (\ref{z})) of the Subsection 5.1. By this reason, the case $V_2=0$ was considered separately in that Subsection.

As for the wave functions for the partner potential $V_1(\vec x),$ they are calculated according to (\ref{NNN}), and the explicit expressions are different depending on the chosen version of $g_1(x_1)$ (Solutions I, \, II or III) and of corresponding solution
$g_2(x_2):$
\ba
\Psi^{(1)}_1(\vec x)&=&\frac{C}{m}[imq_1+m(\partial_1-i\partial_2)g(\vec x)-
2(iq_1-q_2)f(\vec x)]\exp(i\vec q \vec x); \label{q}\\
\Psi^{(1)}_2(\vec x)&=&\frac{C}{m}[-imf(\vec x)
-2i(-iq_1+q_2)(iq_1+(\partial_1+i\partial_2)g(\vec x))]\exp(i\vec q \vec x). \label{qq}
\ea

Again, like in the previous case solutions of $D_1(\vec x)\Omega (\vec x)=0$ which are simultaneously zero modes of operator $M^{\dagger}(\vec x)$ must be built. The latter fact provides the relation for the second component $\Omega_2(\vec x):$
\be
\Omega_2(\vec x)=-if^{-1}(\vec x)[-\partial_1 +(\partial_1+i\partial_2)g(\vec x)]\Omega_1(\vec x),
\label{m-1}
\ee
and again, the second order differential equation for the first component:
\be
\partial_1^2\Omega_1(\vec x)-4(\partial_1g(\vec x))\partial_1\Omega_1(\vec x)+[f^2(\vec x)+3(\partial_1g(\vec x))^2+(\partial_2g(\vec x))^2-(\partial_1^2g(\vec x))]\Omega_1(\vec x)=0.
\label{m-2}
\ee
It is possible to remove the first derivative by:
\be
\Omega_1(\vec x)\equiv \omega_1(\vec x)\exp(2g(\vec x))
\label{m-3}
\ee
so that:
\be
\partial_1^2\omega_1(\vec x)+U(\vec x)\omega_1(\vec x)=0;\quad
U(\vec x)\equiv [f^2(\vec x)-(\partial_1g(\vec x))^2+(\partial_2g(\vec x))^2+(\partial_1^2g(\vec x))]
\label{m-4}
\ee
One can check that the coefficient function $U(\vec x)$ is a constant, whose value depends on a solution number:
\be
U_{I}=\lambda^2/4;\quad U_{II}=-\lambda^2/4;\quad U_{III}=0
\label{m-5}
\ee
For the Solution I,
\be
\Omega^{I}_1(\vec x)=f(\vec x)[f_2(x_2)e^{+i\lambda x_1/2}+
k_2(x_2)e^{-i\lambda x_1/2}],
\label{m-6}
\ee
with arbitrary functions $f_2,\, k_2,$ and the second component $\Omega^{I}_2(\vec x)$ calculated according to (\ref{m-1}).
Arbitrary functions $f_2,\, k_2$ are defined by the second equation $D_1(\vec x)\Omega (\vec x)=0.$ This calculation is straightforward, and we will skip its details.

The result for the Solution I is the following. For the model $V_1^{\alpha},$ besides wave functions (\ref{q}), (\ref{qq}) wave functions of the zero mode kind $\Omega^{I, \alpha}$ exist with components:
\ba
\Omega^{I, \alpha}_1(\vec x)&=&\frac{1}{g_1(x_1)+g_2(x_2)}
[\frac{\lambda +ik}{m}e^{ikx_2/2}e^{i\lambda x_1/2}+\frac{ik}{a}(A^2e^{ikx_2/2}+\frac{b(\lambda +ik)}{m\lambda}e^{-ikx_2/2})e^{-i\lambda x_1/2}];
\label{m-7}\\
\Omega^{I, \alpha}_2(\vec x)&=&\frac{1}{g_1(x_1)+g_2(x_2)}
[e^{ikx_2/2}e^{i\lambda x_1/2}-\frac{ik}{a}(\frac{A^2(\lambda +ik)}{m}e^{ikx_2/2}+\frac{b}{\lambda}e^{-ikx_2/2})e^{-i\lambda x_1/2}];
\label{m-8}
\ea
and the similar zero mode $\Omega^{I, \alpha}$ with $k \to -k.$ In these expressions, all constants
were defined above in this Subsection. Analogously, there are two solutions $\Omega^{I, \beta}$ for the model $V_1^{\beta},$ which are obtained from
(\ref{m-7}), (\ref{m-8}) by $k \to ik.$ For the same particular case with constants described after Eq.(\ref{V-ii}), the zero modes $\Omega^{I, \beta}$
coincide with zero modes (41) in \cite{jakubsky} after constructing the suitable linear combinations. Thus, both kinds of wave functions for the model $V_1^{\beta}$ (see (\ref{q}), (\ref{qq}) and zero modes $\Omega^{I, \beta}$) coincide with those of \cite{jakubsky} in particular case of constants in our model $V_1^{\beta}.$ By suitable transformation, the zero modes of $\beta$-model become the zero modes of the $B$-model in \cite{jakubsky}.

Two zero modes $\Omega^{II}(\vec x)$ for the Solution II are obtained from (\ref{m-7}), (\ref{m-8}) by replacing $\lambda \to i\lambda.$

The zero mode $\Omega^{III}(\vec x)$ for Solution III with $\lambda =0$ can be calculated analogously:
\ba
\Omega^{III}_1(\vec x)&=&\frac{1}{g_1(x_1)+g_2(x_2)}
[(\alpha_1(x_1-1/m)+\frac{mB\beta_1}{c})e^{imx_2/2}+(\beta_1(x_1+1/m)-\frac{mB\alpha_1}{c})e^{-imx_2/2}];
\label{m-9}\\
\Omega^{III}_2(\vec x)&=&\frac{-i}{g_1(x_1)+g_2(x_2)}
[(\alpha_1(x_1+1/m)+\frac{mB\beta_1}{c})e^{imx_2/2}+(\beta_1(x_1-1/m)-\frac{mB\alpha_1}{c})e^{-imx_2/2}];
\label{m-10}
\ea

It was noticed in \cite{jakubsky}, that the wave functions for their models obey the interesting physical property - Super-Klein tunneling, i.e. omnidirectional perfect (reflectionless) transmission through the electrostatic barrier. In the present Subsection, the technical reasons for this property are clear. Due to intertwining, the wave functions $\Psi^{(1)}$ are obtained from an ordinary plane waves with "mass equal $m/2$" $\vec q^2=m^2/4$ by the action of intertwining operator $N(\vec x)$ (see Eqs.(\ref{q}), (\ref{qq})). Due to asymptotic behaviour of hyperbolic cosine in (\ref{ii}), just in the $\beta$-model operator $N(\vec x)$ does not deform asymptotical behaviour of plane waves $\Psi^{(2)}(\vec x).$ Analogously, the system with $g_2(x_2)$ from (\ref{add}) also obeys this property but now due to behaviour of $g_1(x_1)$ from (\ref{6.3}). From other side, obtained explicit forms of solutions $\Psi^{(1)}(\vec x)$ for other models with $g_2^{\alpha}(x_2)$ and $g_2(x_2)$ do not demonstrate such omnidirectional reflectionless transmission of plane waves. As for zero mode solutions above, their behaviour for $|\vec x|\to\infty$ is not of a pure plane wave form.

\subsection{Models with nonconstant $V_2(x_2)$ }

In the previous Subsections, the full solution of Dirac equation for potentials $V_1(\vec x)$ with constant partner potentials $V_2(x_2)$ was derived. It would be interesting to extend the class of models which allow to build both partner potentials $V_2(x_2), \, V_1(\vec x)$ so that the corresponding wave functions could be found. Nonlinearity of the general form Eq.(\ref{3.444}) is the main technical obstacle along this way. This problem can be overcame by suitable choices of $g_2(x_2)$ which originally were arbitrary functions. In the present Subsection, some variants with non-periodic $g_2(x_2)$ expressed in terms of hyperbolic functions will be considered leading to interesting expressions for $V_2(x_2)\neq const$ and for its partner potential $V_1(\vec x).$

\subsubsection{ The case $V_2(x_2)=\mp\frac{\omega}{\cosh(\omega x_2)}$}

For the first such model in the framework of Solution II with $g_1(x_1)=a \cosh(\lambda x_1)$, we take function $g_2(x_2)$ which leads to a full square in the r.h.s. of (\ref{7.111}) simplifying essentially the expression for $\varphi (x_2)$ and, correspondingly, for potential $V_2(x_2):$
\be
g_2(x_2)\equiv \frac{\alpha}{\cosh(\omega x_2)},
\label{ch-1}
\ee
where $\alpha, \omega$ are constants.
By choosing
\be
\alpha \equiv \frac{2\lambda a \omega}{\omega^2+\lambda^2}; \nonumber
\ee
expression for $\varphi$ becomes simple, without square root:
\be
\varphi(x_2)=\pm\frac{\lambda a}{2}\cdot (\frac{2\omega^2}{\omega^2+\lambda^2}\frac{1}{\cosh^2(\omega x_2)}-1).
\label{ch-3}
\ee
Then, the potential $V_2$ has also very compact form with coupling constant coinciding with $\omega :$
\be
V_2(x_2)=\mp\frac{\omega}{\cosh(\omega x_2)},
\label{ch-4}
\ee
and vanishing for $x_2\to\pm\infty.$

This potential (with arbitrary coupling constant in the nominator) appeared in literature in the context of massless Dirac equation with one-dimensional scalar potential $V(x)$ \cite{ho-1}. Although our main task here is the problem with non-trivial partner  potentials $V_1(\vec x)$ depending on both coordinates, those results are useful in order to solve the initial problem with one-dimensional potential $V_2(x_2).$ It was studied in \cite{ho-1} by transformation of Dirac equation to a pair of one-dimensional Schr\"odinger-like second order equations with their own potentials
\be
V_{Schr \pm}(x)=-V^2(x)\pm iV^{\prime}(x),
\label{schr}
\ee
which have a typical supersymmetrical form \cite{reviews} -- \cite{reviews-5} but with pure imaginary superpotential $iV(x).$ The Schr\"odinger equations with potentials (\ref{schr}) define two solutions $\eta_+, \eta_-,$ correspondingly. These solutions are related to the components of solution $\Psi^{(2)}$ of Dirac equation as follows:
\be
\Psi^{(2)}_{1}(\vec x)\equiv \exp(ik_1x_1)(\eta_+(x_2)+\eta_-(x_2));\,
\Psi^{(2)}_{2}(\vec x)\equiv \exp(ik_1x_1)(\eta_+(x_2)-\eta_-(x_2)).
\label{ch-5}
\ee
Here, due to independence of $V_2$ on $x_1,$ both components of $\Psi^{(2)}$ include the plane wave along $x_1$ with momentum value $k_1,$ while functions $\eta_+, \eta_-$ are solutions of the Schr\"odinger equation with potentials (\ref{schr}) and energy $-k_1^2.$

The well developed methods of SUSY Quantum Mechanics provide the exact solution of the Schr\"odinger problem (\ref{schr}) with a variety of superpotentials. If we take $V(x)$ in the r.h.s. of (\ref{schr}) coinciding with our $V_2(x)$ from (\ref{ch-4}), the obtained potentials (\ref{schr}) are the well known Scarf II \cite{scarf} potentials:
\be
V_{Schr}(x)=\frac{B^2-A^2-A\omega}{\cosh^2(\omega x)}
\pm\frac{B(2A+\omega)\sinh(\omega x)}{\cosh^2(\omega x)}
,
\label{scarf}
\ee
$A, B$ are constants.
The Scarf II potential obeys the property of shape invariance \cite{shape} and therefore, it belongs to the class of exactly solvable potentials \cite{dabrowska}. Thus, we may use these results (e.g. see \cite{IKN-2}, \cite{IKN-1}) and find the components $\eta_{\pm}$ in terms of Jacobi polynomials \cite{erd} as follows.
\ba
\eta_{n, +}(y)&=&i^n(1+y^2)^{\frac{1}{4}(\gamma + \beta+1)}\exp(-\frac{i}{2}(\gamma -\beta)\arctan(y))P_n^{(\gamma ,\beta)}(iy);
\label{ch-6}\\
\eta_{n,-}(y)&=&i^n(1+y^2)^{\frac{1}{4}(\gamma + \beta+1)}\exp(+\frac{i}{2}(\gamma -\beta)\arctan(y))P_n^{(\beta ,\gamma)}(iy),
\label{ch-7}
\ea
where $y\equiv \sinh(\omega x_2), \,\,n$ is an integer numerating energy eigenvalues $k_1^2$ and
\be
\beta\equiv -\frac{A}{\omega}+\frac{iB}{\omega}-\frac{1}{2};\quad
\gamma\equiv -\frac{A}{\omega}-\frac{iB}{\omega}-\frac{1}{2};\quad k_1^2=(A-n\omega)^2.
\label{ch-8}
\ee
There are four options for the pairs $(\gamma, \beta)$ in our case with specific potential (\ref{ch-4}):
\be
(\gamma, \beta) = (\pm \frac{1}{2}, \,\pm \frac{3}{2}),
\label{ch-9}
\ee
with arbitrary combination of signs. One of these pairs $(-1/2,\, -3/2)$ satisfies the condition of normalizability of $\eta_{\pm}(x_2)$ (see \cite{dabrowska}, \cite{IKN-2}, \cite{IKN-1}, \cite{ho-1}) $A = - \frac{1}{2}(\beta + \gamma +1)\omega > 0,$ providing the normalizable in $x_2$-direction wave function $\Psi^{(2)},$ which has not the form of plane wave in asymptotics. Nevertheless, the whole two-dimensional wave function $\Psi^{(2)}(\vec x)$ is not normalizable in a plane due to periodic  exponents in expressions (\ref{ch-5}).

The partner potential is calculated according to (\ref{4.1}) depending also on function $g_1(x_1)=a\cosh(\lambda x_1):$
\be
V_1(\vec x)=V_2(x_2)+\frac{2\varphi (x_2)}{g_1(x_1)+g_2(x_2)}=\mp\frac{\omega}{\cosh(\omega x_2)}\pm \frac{\lambda a\cdot (\frac{2\omega^2}{\omega^2+\lambda^2}\frac{1}{\cosh^2(\omega x_2)}-1)}{a \cosh(\lambda x_1)+\frac{2\lambda a\omega}{(\omega^2+\lambda^2)\cosh(\omega x_2)}}.
\label{VVV}
\ee

This potential has the form of a surface with two mutually orthogonal valleys (along $x_1$ and $x_2$), the depths of them depend on the parameters of the potential. For illustration, Figure 4 shows the shape of the potential with different values of parameters $\omega,\, \lambda,$ and $a.$
The wave functions $\Psi^{(1)}(\vec x)$ can be built according to (\ref{NNN}) by action of operator $N(\vec x),$ which gives also non-normalizable wave functions $\Psi^{(1)}(\vec x).$.

\subsubsection{The case $V_2(x_2)=\mp\frac{\lambda c}{4\cosh(\frac{\lambda x_2}{2})}.$}

Let us take another special form of $g_2(x_2)$ which also simplifies essentially expression for $\varphi (x_2)$ but now in the framework of Solution I with $g_1(x_1)=a \cos(\lambda x_1):$
\be
g_2(x_2)=b\cosh(\lambda x_2)+bc^2  \quad b^2(1-c^2)=a^2;\quad |c| < 1.
\label{33-1}
\ee
It leads to
\be
\varphi(x_2)=\pm\lambda bc\cosh(\frac{\lambda x_2}{2}), \label{33-2}
\ee
and the potential:
\be
V_2(x_2)=\mp\frac{\lambda c}{4\cosh(\frac{\lambda x_2}{2})}
\label{u-1}
\ee
with arbitrary constant $|c|<1.$ Actually, Dirac equation with such one-dimensional potential (up to a scale of coordinate) was considered by SUSY methods in the paper \cite{ho-1}, already mentioned above (see also \cite{IKN-2}, \cite{IKN-1} on Scarf II potential). The wave functions $\Psi^{(2)}(\vec x)$ are expressed also in terms of Jacobi polynomials $P_n^{(\gamma, \beta)}$ analogously to Eqs.(\ref{ch-5}) -- (\ref{ch-8}) but with four possible pairs of indices $(\gamma ,\, \beta):$
\be
(\beta,\,\gamma) = (\pm\frac{c+1}{2},\, \pm\frac{c-1}{2});  \quad  |c| < 1
\label{u-2}
\ee
with arbitrary combinations of signs. For $|c| < 1$, all these options lies beyond the condition of normalizability for $\Psi^{(2)}(x_2).$

The potential $V_1(\vec x)$ is:
\be
V_1(\vec x)=\mp\frac{\lambda c}{4\cosh(\frac{\lambda x_2}{2})}\pm
\frac{2\lambda bc\cosh(\frac{\lambda x_2}{2})}{a\cos(\lambda x_1)+b(\cosh(\lambda x_2)+c^2)},
\label{33-3}
\ee
the dependence of its form and scale on changing parameters $c,$ and $\lambda$ is demonstrated at Figure 5. The wave functions $\Psi^{(1)}(\vec x)=N(\vec x)\Psi^{(2)}(x_2)$ are built by means of intertwining operator (\ref{NNN}).

\subsubsection{New potential.}

Let us start a new model by taking Solution I with exponential form for the function $g_2:$
\be
g_2(x_2)=\frac{b}{2}\exp(\lambda x_2)+c,
\label{e-1}
\ee
with arbitrary real constants $b, c, \lambda.$ The corresponding function $\varphi$ is:
\be
\varphi(x_2)=\frac{\lambda}{2}\sqrt{bc\exp(\lambda x_2)+c^2-a^2}.
\label{e-2}
\ee
Then, expressions for potential $V_2(x_2)$ and its partner $V_1(\vec x)$ are:
\ba
V_2(x_2)&=&-\frac{\lambda c}{2\sqrt{bc\exp(\lambda x_2)+c^2-a^2}};
\label{e-3}\\
V_1(\vec x)&=&V_2(x_2)+
\frac{\lambda \sqrt{bc\exp(\lambda x_2)+c^2-a^2}}{a\cos(\lambda x_1)+\frac{b}{2}\exp(\lambda x_2)+c}.
\label{e-4}
\ea
The form of the latter potential is illustrated by Figure 6 for different values of parameters $c\,$ and $ \lambda, \, $ of the model.

To solve the Dirac equation $D_2\Psi^{(2)}=0$ one can transform it to the form of one-dimensional Schr\"odinger equation for the first component $\Psi^{(2)}_1(x_2):$
\be
(\Psi^{(2)}_1)^{\prime\prime} -\frac{V_2^{\prime}}{V_2}(\Psi^{(2)}_1)^{\prime}+(V_2^2-k_1\frac{V_2^{\prime}}{V_2}-k_1^2)\Psi^{(2)}_1=0;\quad \Psi^{(2)}_2=-V_2^{-1}((\Psi^{(2)}_1)^{\prime}+k_1\Psi^{(2)}_1).
\label{e-5}
\ee
This second order equation can be rewritten \cite{polyanin} in the form of hypergeometric equation:
\be
t(t-1)y'' + [(2\omega + \frac{3}{2})t-(2\omega +1)]y' + [\omega^2 +\frac{\omega}{2}+\frac{k_1(\lambda -2k_1)}{2\lambda^2}]y=0,
\label{e-6}
\ee
where the following definitions and substitutions were made:
\be
y\equiv \Psi_1^{(2)} e^{\lambda \omega x_2}; \, \quad  \omega \equiv \pm[\frac{k_1^2}{\lambda^2}-\frac{c^2}{4(c^2-a^2)}]^{1/2};\quad         t \equiv - \frac{1}{A}e^{\lambda x_2};\quad A \equiv \frac{c^2-a^2}{bc}>0.
\label{e-7}
\ee

\section{Conclusions}

The new form of SUSY-like intertwining relations for two-dimensional massless Dirac operators was used above to study an opportunities to solve analytically Dirac equation with scalar potentials $V_1(\vec x)$ depending on both coordinates.
A few suitable ansatzes for the partner potential $V_2(x_2)$ which is either constant or depends only on one of coordinates, are suitable to solve this problem. As a result, a series of two-dimensional potentials $V_1(\vec x)$ is obtained, each of them includes several free parameters, and all of potentials non-trivially depend on both coordinates. For illustration, some representatives of these potentials are given graphically in Figures 1-6. This approach provides the wave functions not only obtained directly from intertwining $\Psi^{(1)}=N\Psi^{(2)}$ but also the so called zero modes $M^{\dagger}\Omega = 0.$ The specific form of these wave functions will depend both on the chosen model for potential and on the boundary conditions which are dictated by a specific realistic experimental situation (see, for example, \cite{ex}).

\vspace{0.7cm}

{\bf\large{Data Availability Statement:}}

No Data associated in the manuscript.

\newpage
\begin{figure}[t]
\vspace{-3cm}
{\bf\large{Appendix. Figures.}}
\begin{center}
\vspace{-1cm}
\includegraphics[width=1.15\textwidth]{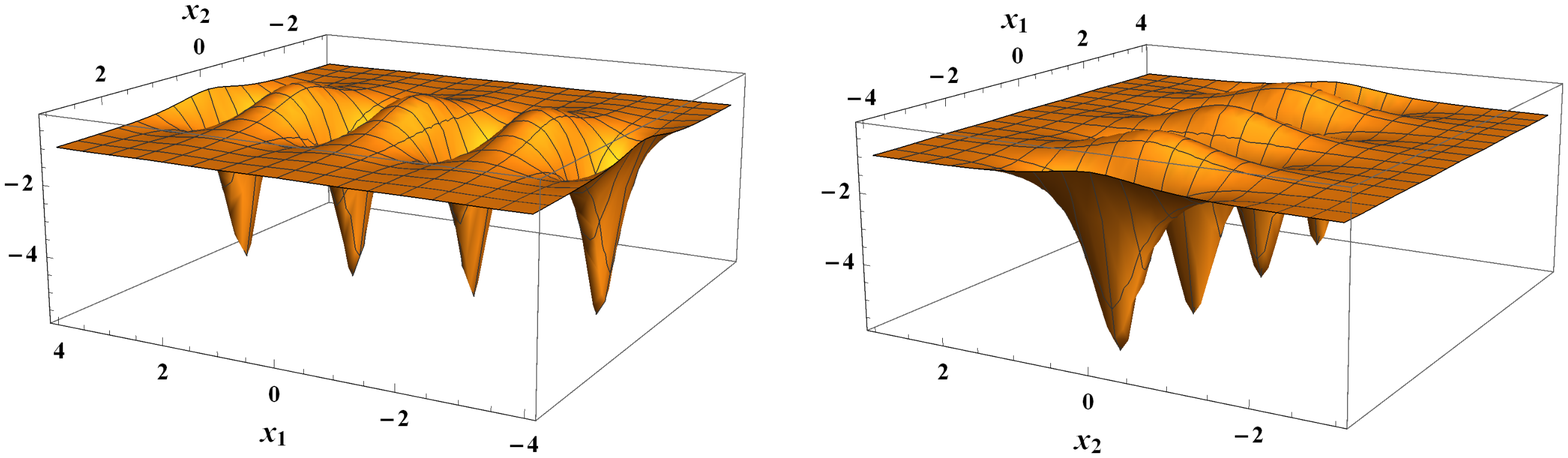}
\vspace{-8.5cm}
\center{a)}
\vspace{-0.40cm}
\includegraphics[width=1.15\textwidth]{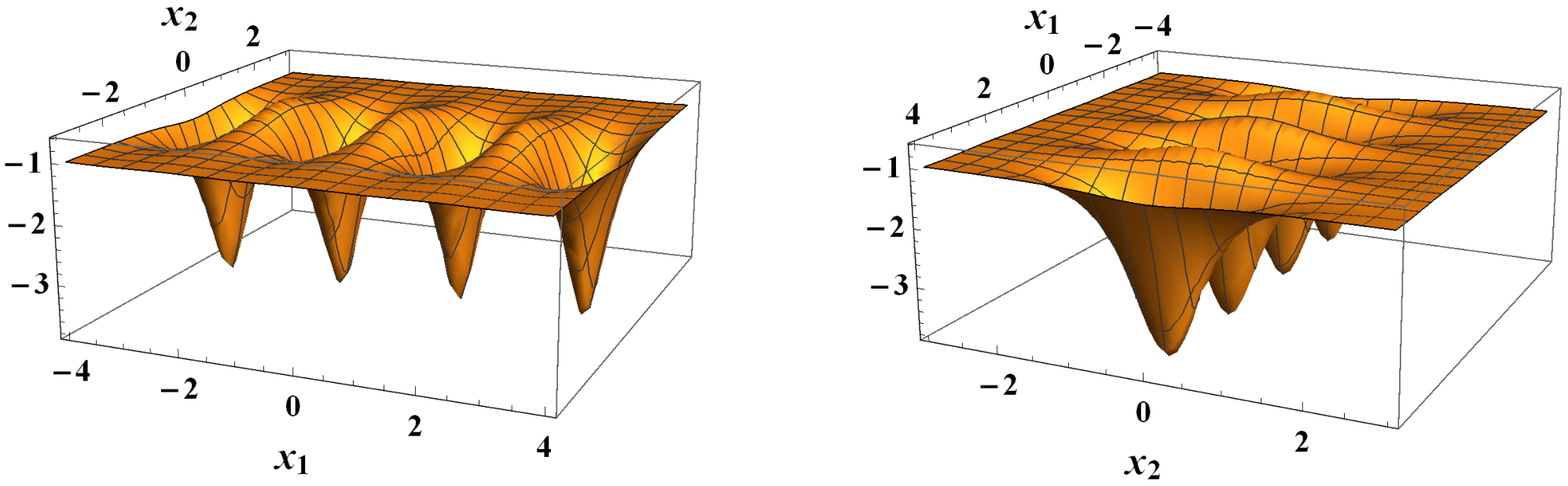}
\vspace{-8.5cm}
\center{b)}
\vspace{-0.40cm}
\includegraphics[width=1.15\textwidth]{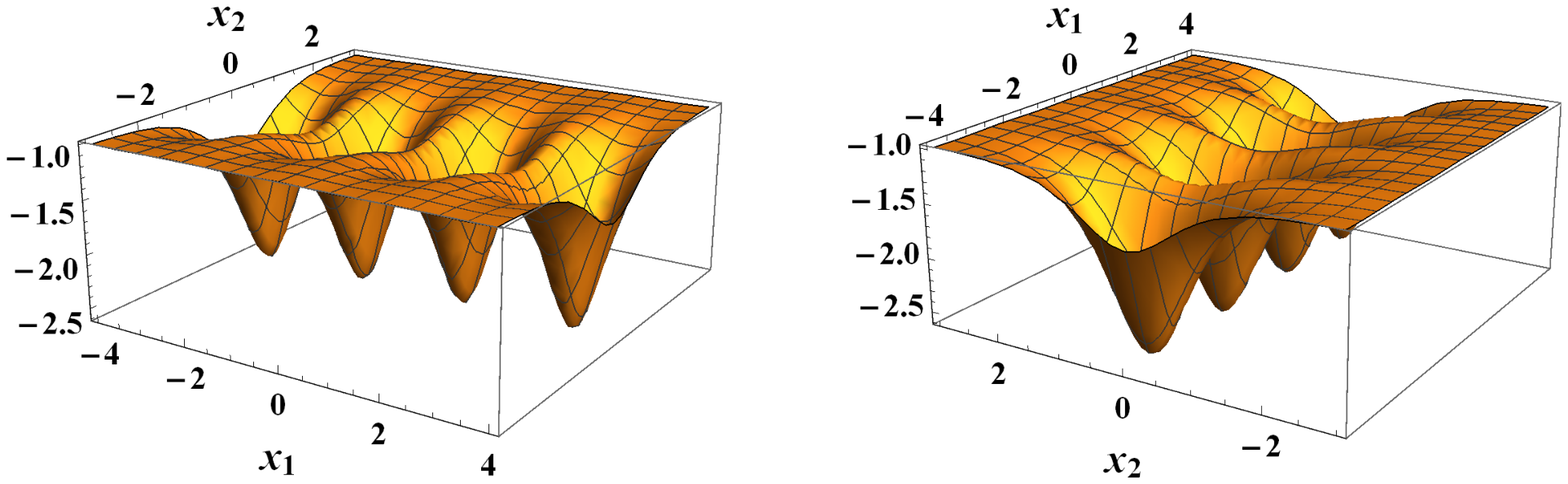}
\vspace{-8.5cm}
\center{c)}
\end{center}

\caption{Potential of Eq.(\ref{V-ii}) with $m=b=k=2$ for different values of $A:\,$ $A=0.0$ (Panel a),\, $A=0.3$ (Panel b), \, $A=0.6$ (Panel c).}
\end{figure}
\newpage
\begin{figure}[t]
\vspace{-3cm}
\begin{center}
\vspace{-1cm}
\includegraphics[width=1.15\textwidth]{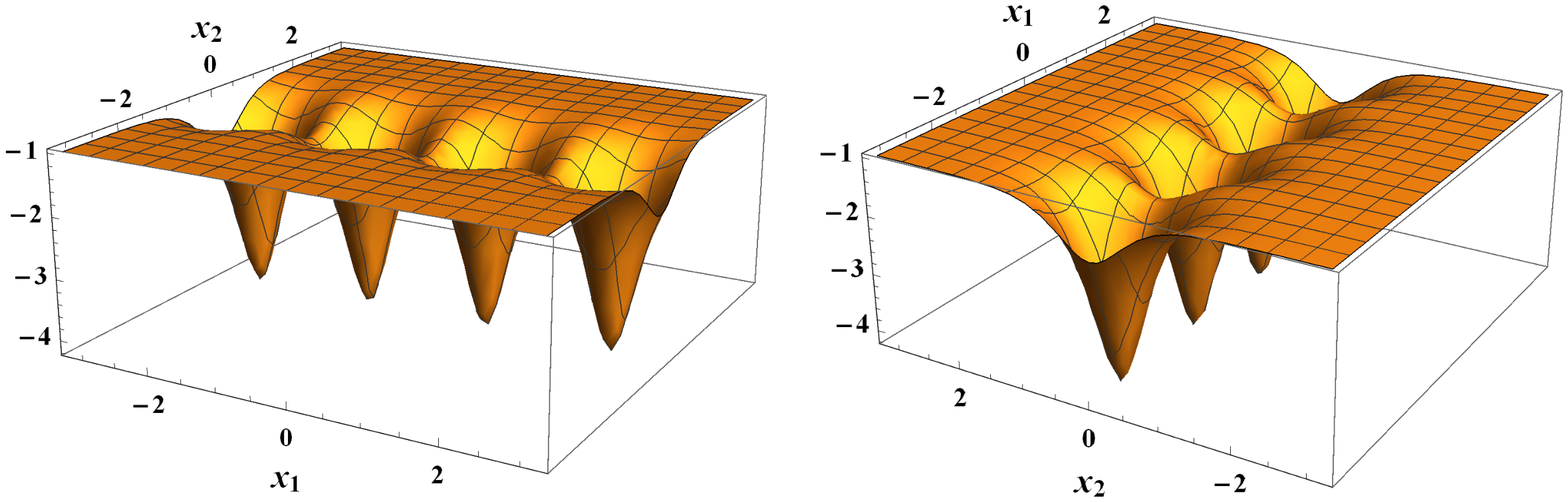}
\vspace{-8.5cm}
\center{a)}
\vspace{-0.40cm}
\includegraphics[width=1.15\textwidth]{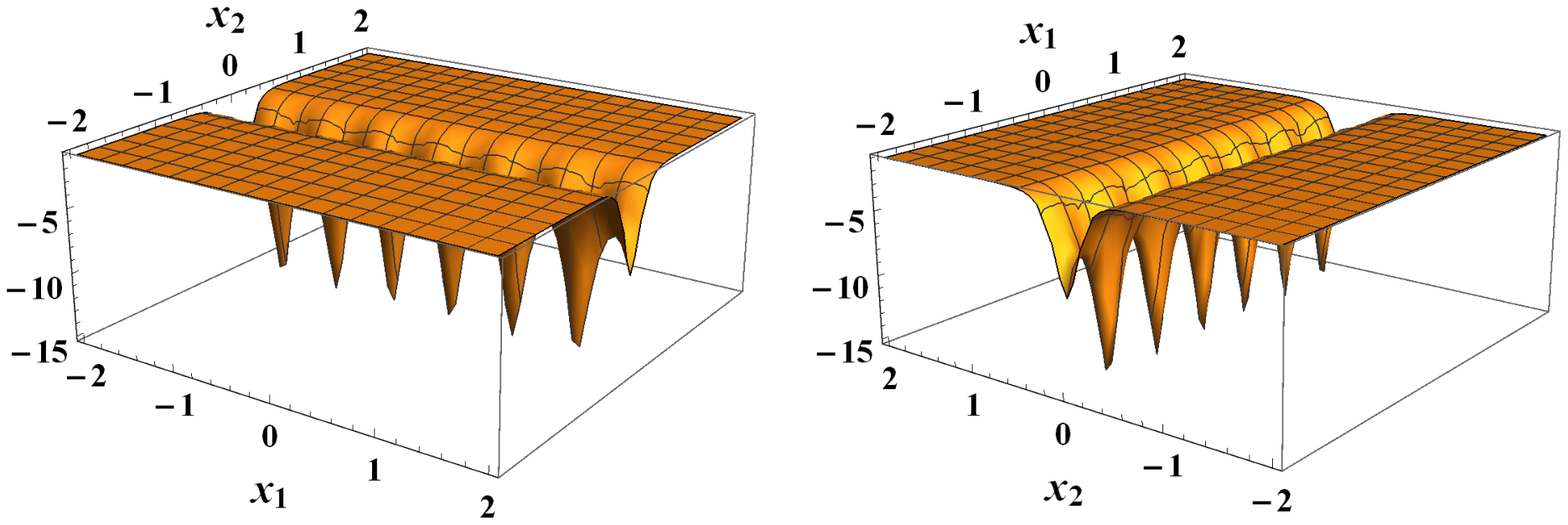}
\vspace{-8.5cm}
\center{b)}
\vspace{-0.40cm}
\includegraphics[width=1.15\textwidth]{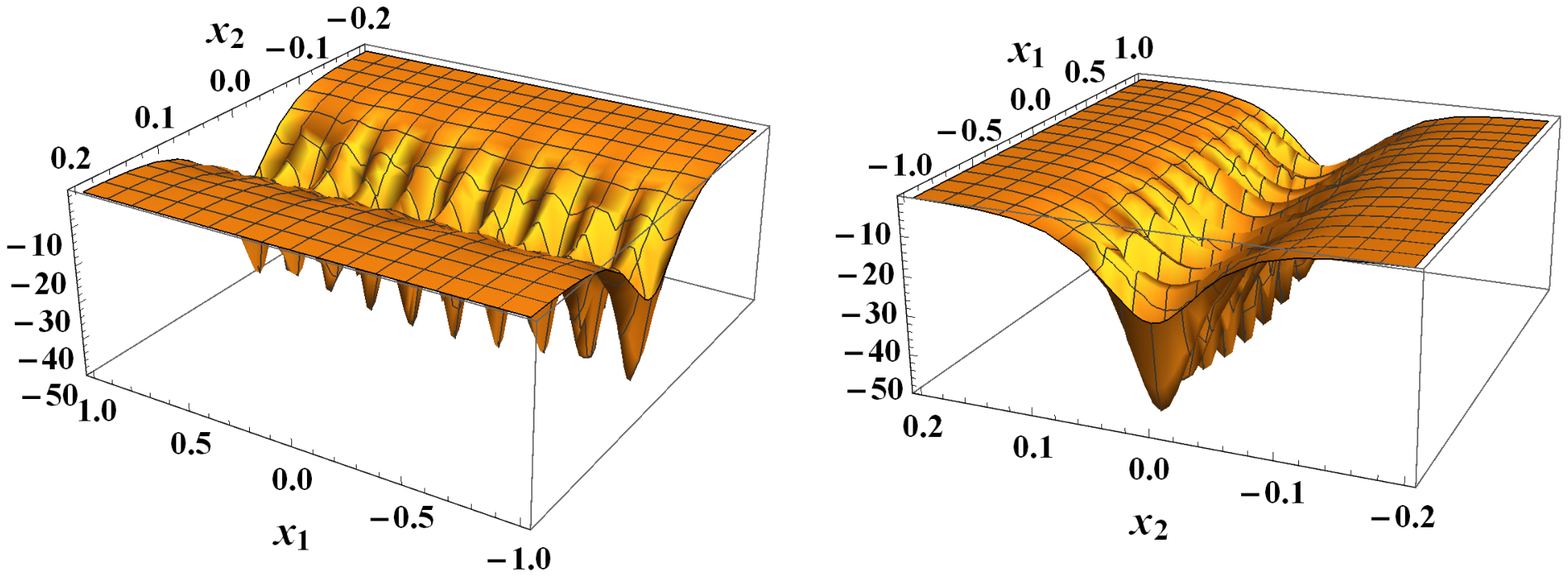}
\vspace{-8.5cm}
\center{c)}
\end{center}

\caption{Potential of Eq.(\ref{V-ii}) with $m=b=2,\, a=1$ for different values of $k:\,$ $k=3$ (Panel a),\, $k=10$ (Panel b), and \, $k=30$ (Panel c).}
\end{figure}

\newpage
\begin{figure}[t]
\vspace{-5cm}
\begin{center}
\vspace{-1cm}
\includegraphics[width=1.15\textwidth]{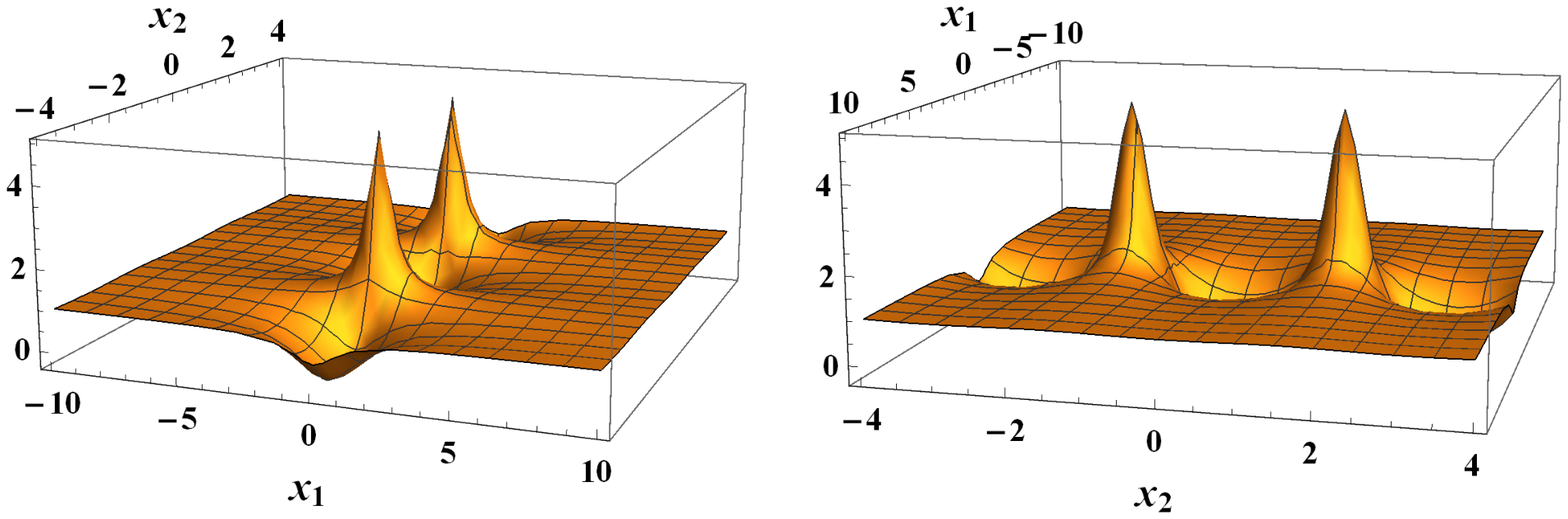}
\vspace{-7cm}
\center{a)}
\vspace{-2cm}
\includegraphics[width=1.15\textwidth]{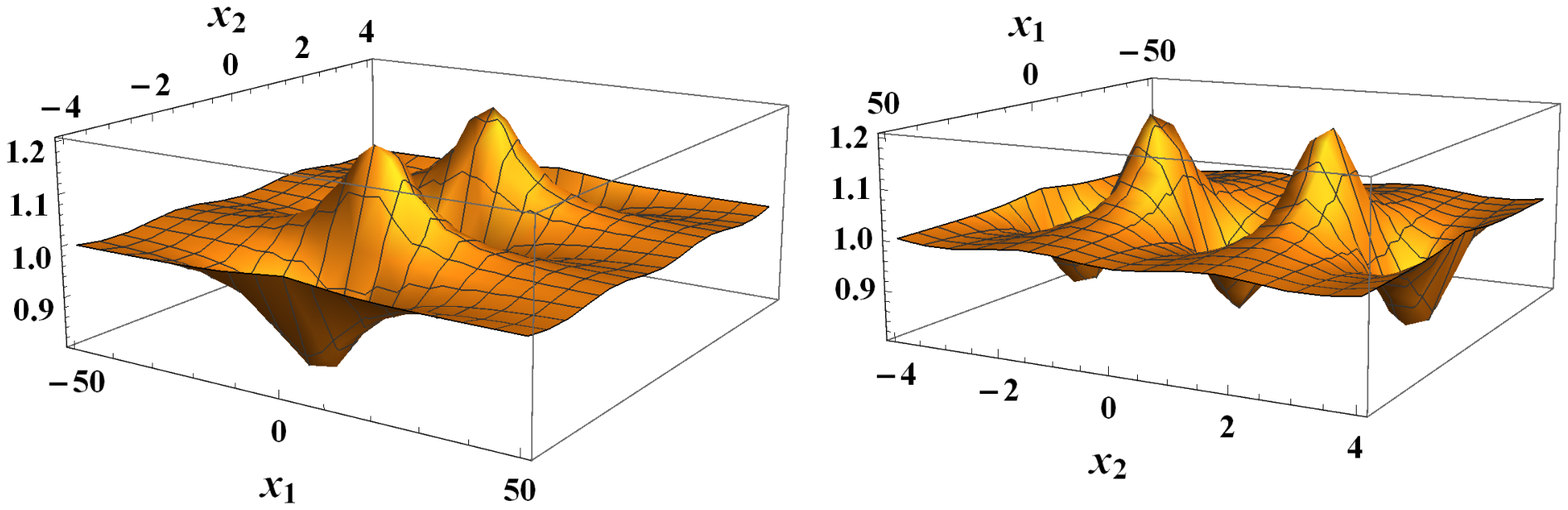}
\vspace{-7cm}
\center{b)}
\end{center}
\end{figure}

\newpage
\begin{figure}[t]
\vspace{-5cm}
\begin{center}
\vspace{-1cm}
\includegraphics[width=1.15\textwidth]{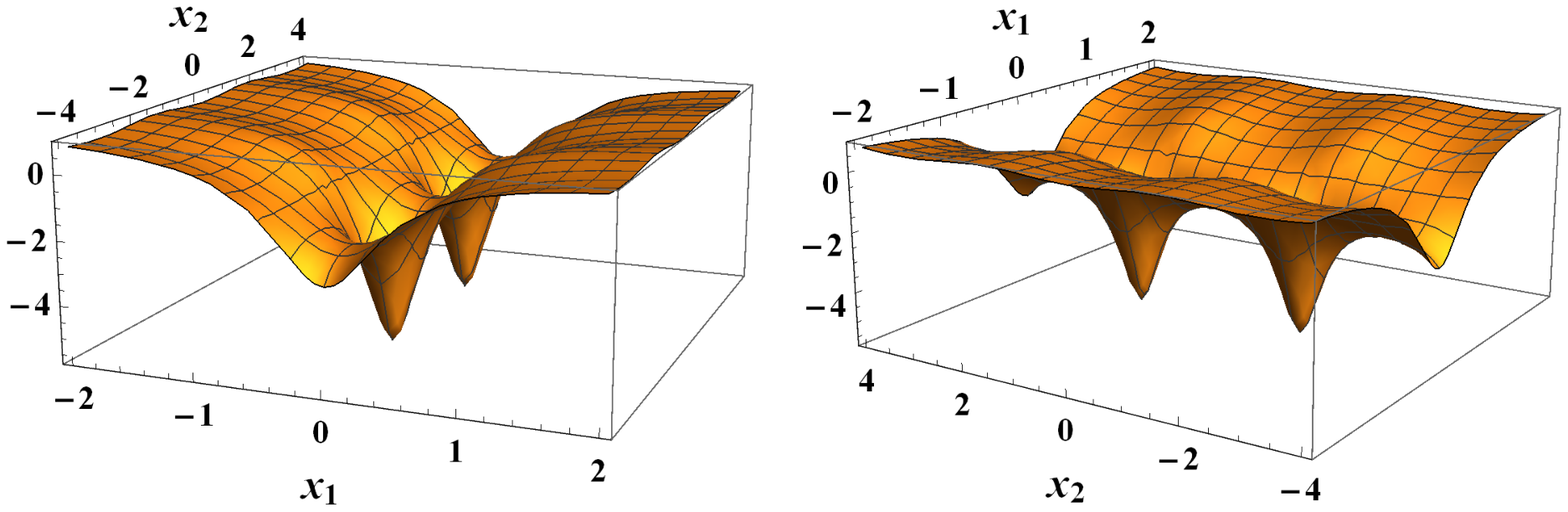}
\vspace{-7cm}
\center{c)}
\vspace{-2cm}
\includegraphics[width=1.15\textwidth]{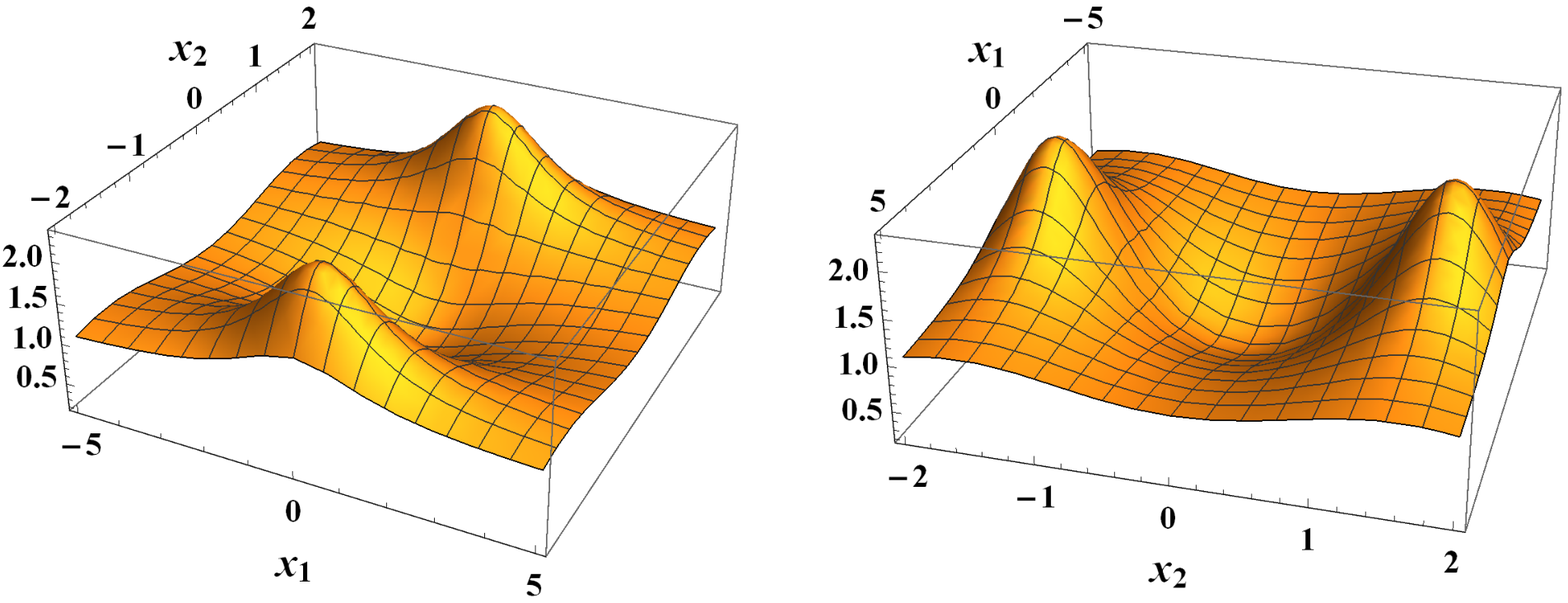}
\vspace{-7cm}
\center{d)}
\end{center}

\caption{Potential of Eq.(\ref{V-i-3}) with $m=2$ for $c=2,\, B=1$ (Panel a), for $c=2, \, B=10$ (Panel b), for $c=10, \, B=1 $ (Panel c), and for $c=10, \, B=10$ (Panel d).}
\end{figure}

\newpage
\begin{figure}[t]
\vspace{-5cm}
\begin{center}
\vspace{-1cm}
\includegraphics[width=1.15\textwidth]{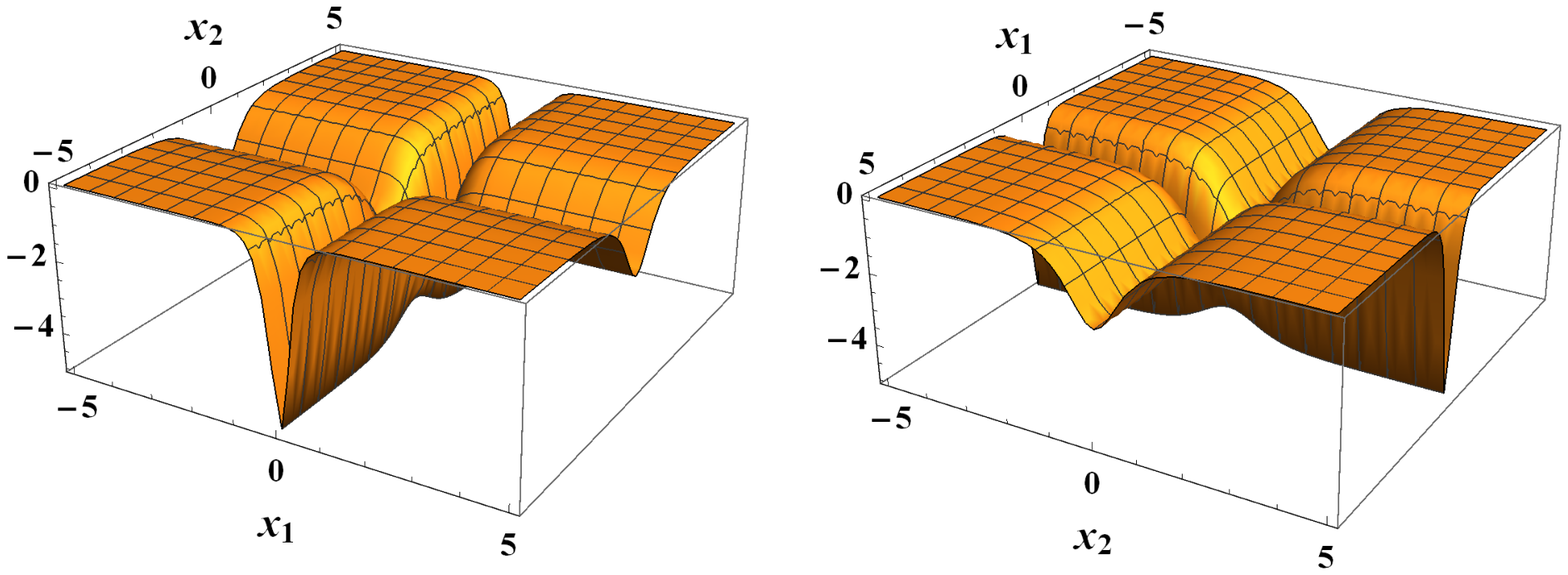}
\vspace{-7cm}
\center{a)}
\vspace{-2cm}
\includegraphics[width=1.15\textwidth]{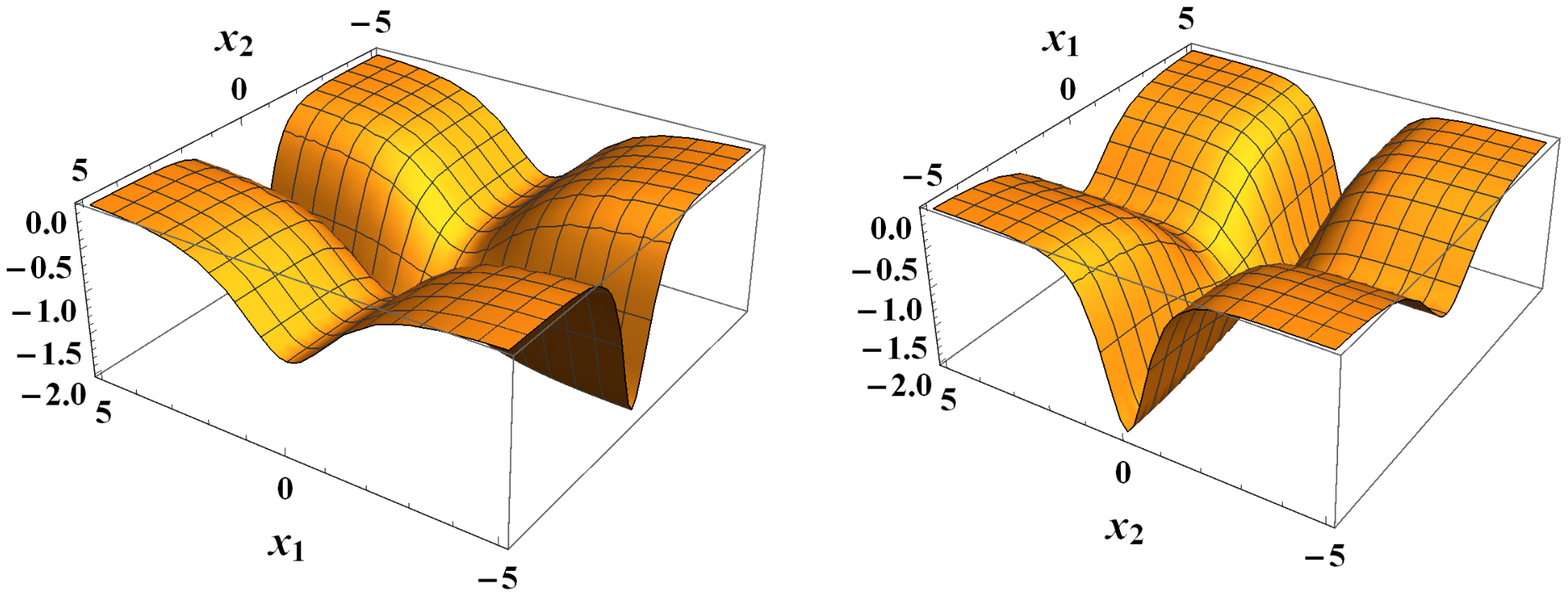}
\vspace{-7cm}
\center{b)}
\end{center}
\end{figure}

\newpage
\begin{figure}[t]
\vspace{-5cm}
\begin{center}
\vspace{-1cm}
\includegraphics[width=1.15\textwidth]{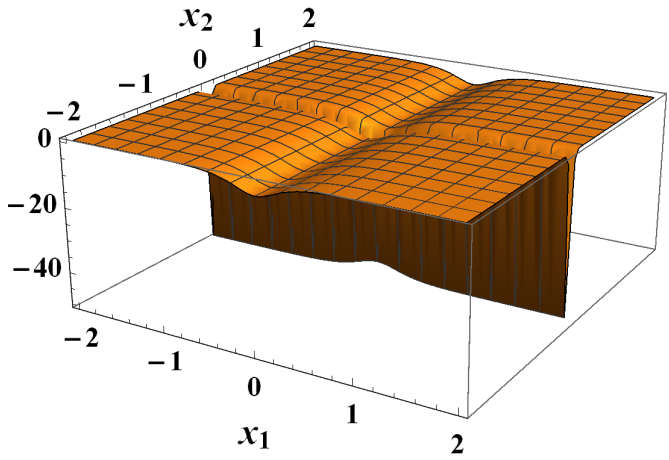}
\vspace{-7cm}
\center{c)}
\vspace{-2cm}
\includegraphics[width=1.15\textwidth]{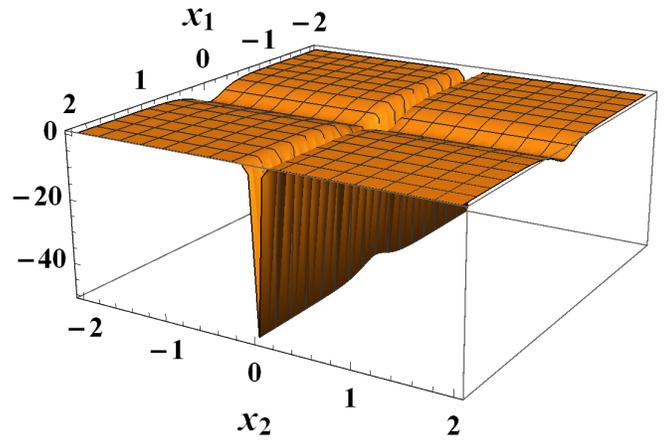}
\vspace{-7cm}
\center{d)}
\end{center}

\caption{Potential of Eq.(\ref{VVV})  for $\omega = 2,\, \lambda = 5, \, a=1$ (Panel a), for $\omega = 2,\, \lambda = 1, \, a=1$ (Panel b), for $\omega = 2,\, \lambda = 1, \, a=100$ (Panel c), and for $\omega = 50,\, \lambda = 5, \, a=1$ (Panel d).}
\end{figure}

\newpage
\begin{figure}[t]
\vspace{-2cm}
\begin{center}
\vspace{-1cm}
\includegraphics[width=1.15\textwidth]{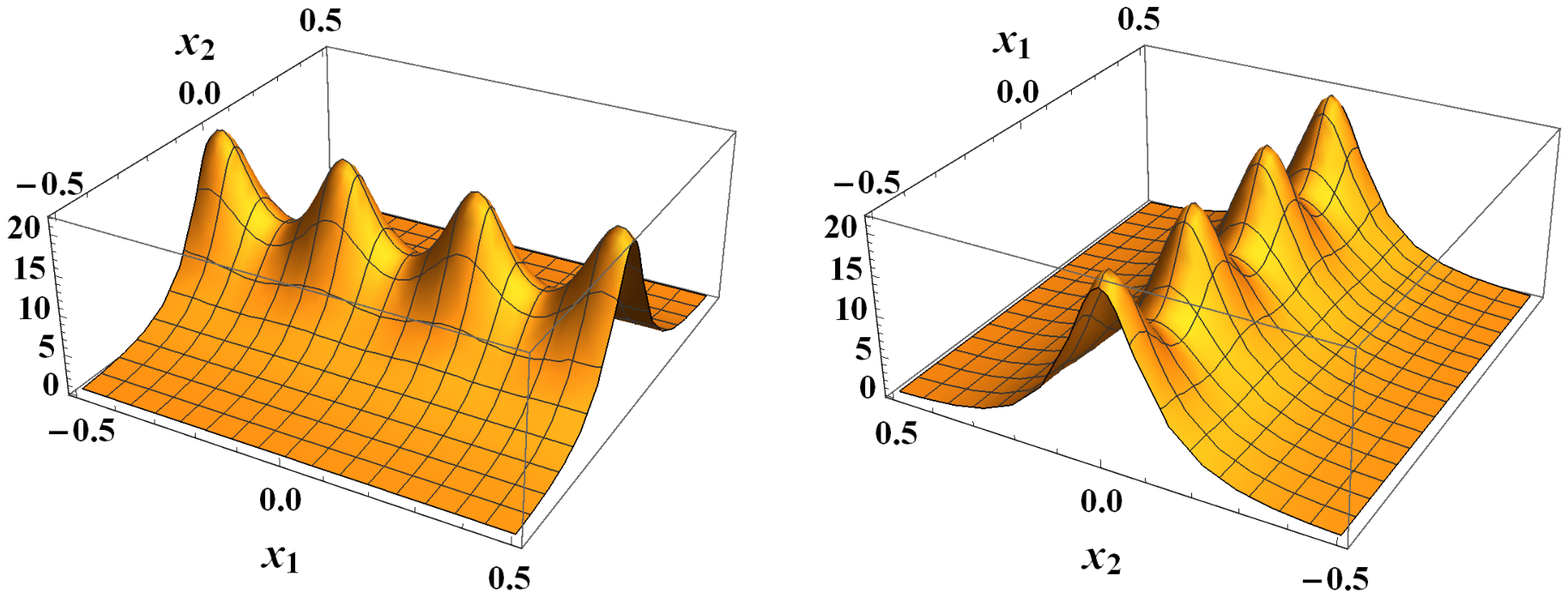}
\vspace{-8.5cm}
\center{a)}
\vspace{-0.40cm}
\includegraphics[width=1.15\textwidth]{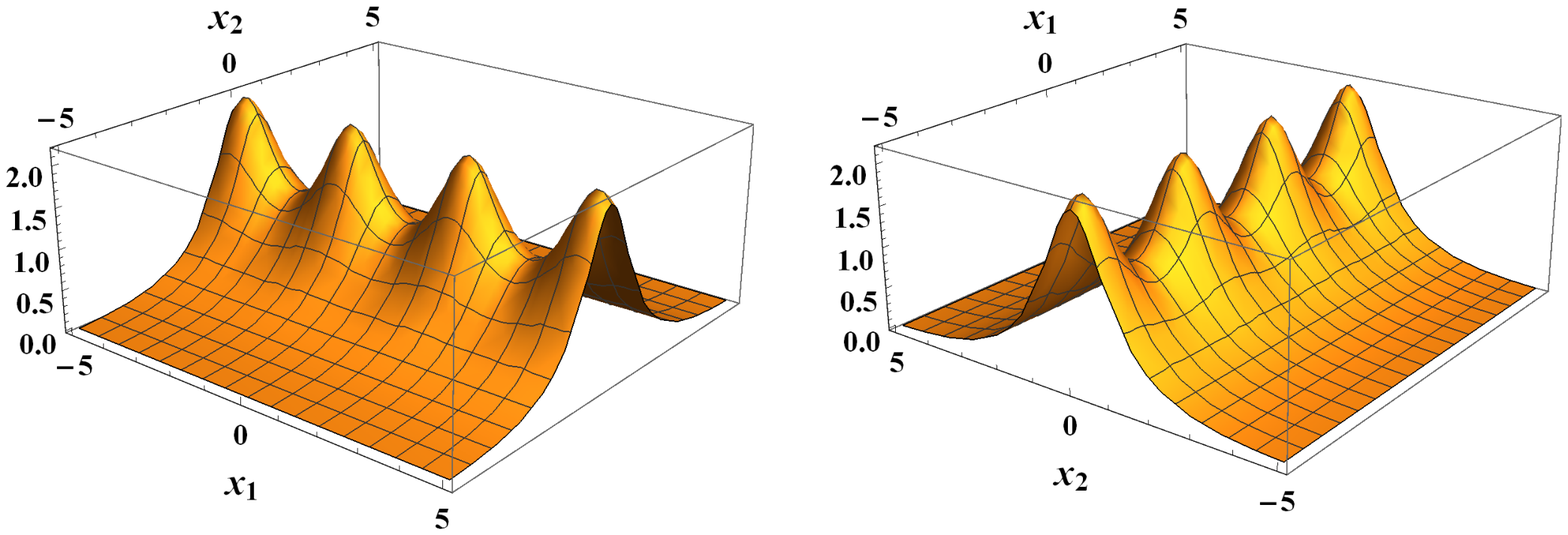}
\vspace{-8.5cm}
\center{b)}
\vspace{-0.40cm}
\includegraphics[width=1.15\textwidth]{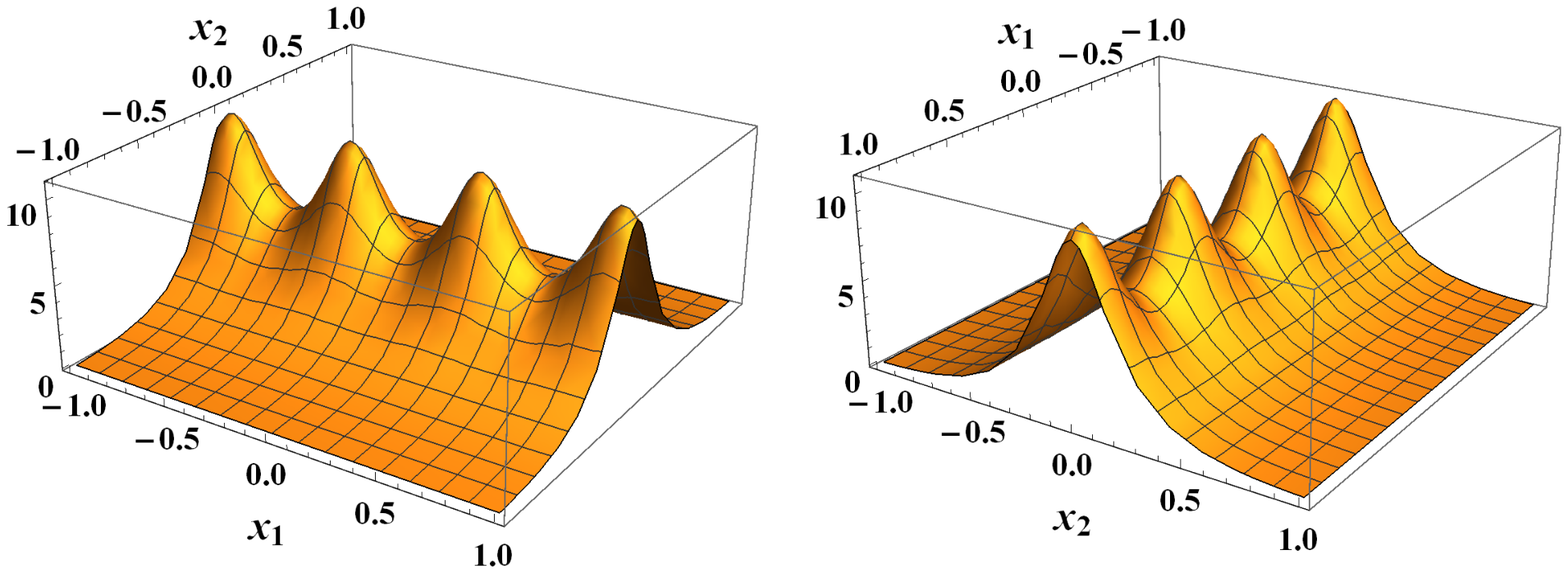}
\vspace{-8.5cm}
\center{c)}
\end{center}

\caption{Potential of Eq.(\ref{33-3}) for $b=10, c = 0.1, \, \lambda = 10$ (Panel a), \, for $b=10, \, c=0.9, \, \lambda = 2$ (Panel b), and for $b=10, \, c=0.9, \, \lambda = 10$ (Panel c).}
\end{figure}

\newpage
\begin{figure}[t]
\vspace{-2cm}
\begin{center}
\vspace{-1cm}
\includegraphics[width=1.15\textwidth]{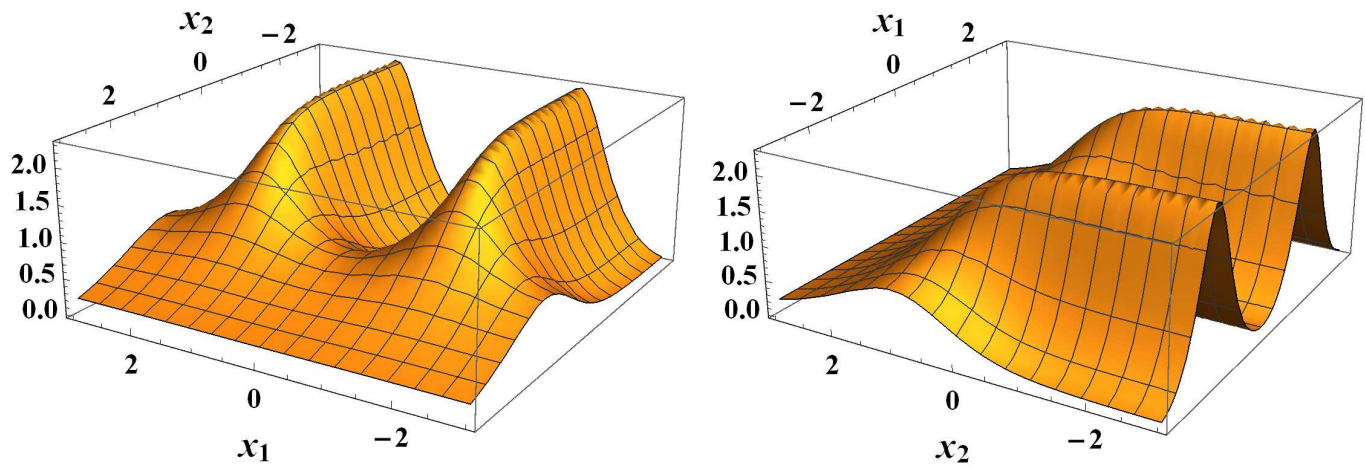}
\vspace{-8.5cm}
\center{a)}
\vspace{-0.40cm}
\includegraphics[width=1.15\textwidth]{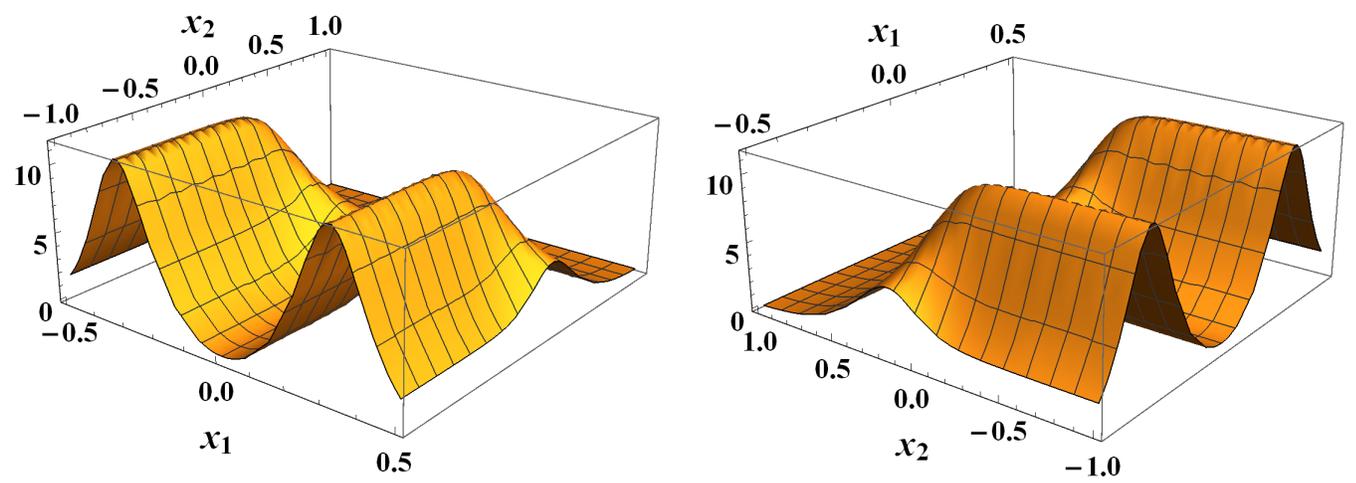}
\vspace{-8.5cm}
\center{b)}
\vspace{-0.40cm}
\includegraphics[width=1.15\textwidth]{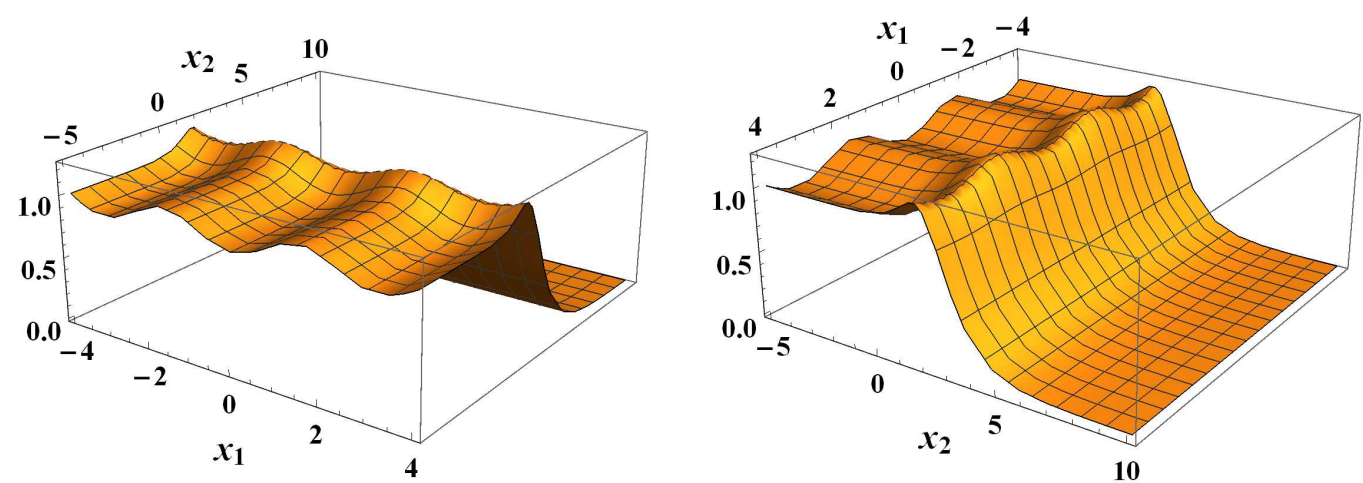}
\vspace{-8.5cm}
\center{c)}
\end{center}

\caption{Potential of Eq.(\ref{e-4}) with $a=b=1$ for $c = \lambda =2$ (Panel a), for $c = 2,\, \lambda = 10$ (Panel b), and for $c = 20, \, \lambda = 2$ (Panel c).}
\end{figure}

\end{document}